\documentclass[preprintnumbers,showkeys,showpacs,amsmath,amssymb]{revtex4}
\usepackage{amsmath,amssymb,graphics,epsfig,subfigure}
\usepackage{color}
\usepackage{multirow}
\usepackage{booktabs}
\usepackage{changes}

\usepackage{hyperref}
\hypersetup{
    colorlinks=true,
    linkcolor=blue,
    citecolor=magenta
}

\begin{document}
\renewcommand{\baselinestretch}{1.3}

\title{Distinct topological configurations of equatorial timelike circular orbit for spherically symmetric (hairy) black holes}

\author{Xu Ye, Shao-Wen Wei \footnote{Corresponding author. E-mail: weishw@lzu.edu.cn}}

\affiliation{$^{1}$Lanzhou Center for Theoretical Physics, Key Laboratory of Theoretical Physics of Gansu Province, School of Physical Science and Technology, Lanzhou University, Lanzhou 730000, People's Republic of China,\\
$^{2}$Institute of Theoretical Physics $\&$ Research Center of Gravitation,
Lanzhou University, Lanzhou 730000, People's Republic of China}

\begin{abstract}
Topology is a promising approach toward to the light ring in a generic black hole background, and equatorial timelike circular orbit in a stationary black hole background. In this paper, we consider the distinct topological configurations of the timelike circular orbits in static, spherically symmetric, and asymptotic flat black holes. By making use of the equation of motion of the massive particles, we construct a vector with its zero points exactly relating with the timelike circular orbits. Since each zero point of the vector can be endowed with a winding number, the topology of the timelike circular orbits is well established. Stable and unstable timelike circular orbits respectively have winding number +1 and -1. In particular, for given angular momentum, the topological number of the timelike circular orbits also vanishes whether they are rotating or not. Moreover, we apply the study to the Schwarzschild, scalarized Einstein-Maxwell, and dyonic black holes, which have three distinct topological configurations, representations of the radius and angular momentum relationship, with one or two pairs timelike circular orbits at most. It is shown that although the existence of scalar hair and quasi-topological term leads to richer topological configurations of the timelike circular orbits, they have no influence on the total topological number. These results indicate that the topological approach indeed provides us a novel way to understand the timelike circular orbits. Significantly, different topological configurations can share the same topology number, and hence belong to the same topological class. More information is expected to be disclosed when other different topological configurations are present.
\end{abstract}


\maketitle

\section{Introduction}\label{sec1}

The gravitational wave detections by LIGO and Virgo Collaborations \cite{ligo16a,ligo16b,ligo17} provide strong evidence that astrophysical black holes exist and merge. Via such binary black hole merger, the nature of the black hole can be well studied with the inspiral, merger, and ringdown waveforms. On the other hand, through the observations of shadow imaging \cite{EHTL1,EHTL5,EHTL6}, the information near the black hole horizon geometry can be tested.

Extensive studies have shown that the ringdown and shadow observables are both intimately connected to a special set of null circular orbits known as the light rings (LRs) \cite{Cardoso2016,Cunha2018}. Apart from the null geodesics, the timelike geodesics of massive particles can also form the circular orbit around the black holes. Such timelike circular orbits (TCOs) are also one kind fundamental characteristic orbit. These massive particles dropped from far away from the black hole accumulate on these stable TCOs and form an accretion disk with its inner edge measuring by the innermost stable circular orbit (ISCO) \cite{Abramowicz2011}.

On account of that these characteristic orbits are directly related to the motion of particles that hide valuable information on spacetime background, several different methods are developed to deal with these circular orbits. The most common one is to solve the geodesic equations via Lagrangian, and obtain the circular orbit by formulating the effective potential. This treatment has a wide range of applications such as studying the photon sphere in static, stationary, or dynamical spacetime \cite{Hasse2001,Koga2022,Johannsen2013}. The other one is called the quasi-local approach, through which the first quasi-local definition of the photon surface was given by Claudel, Virbhadra, and Ellis \cite{Claudel2000}, and then it is extended to the trapped surface \cite{Shiromizu2017}. For a massive particle surface, Kobialko, Bogush and Galtsov established a significant theorem which describes the pure geometry of timelike surface without requiring the worldline dynamics equation \cite{Kobialko2022}.

Quite differently, the topological method can also be applied in the analysis of circular orbit whether its center is a black hole or a horizonless ultracompact object. In Ref. \cite{Cunha2017}, Cunha, Berti, and Herdeiro first proposed a topological approach and proved a theorem that if an axisymmetric and stationary solution of the Einstein field equation obeys the null energy condition, the ultracompact objects formed from the classical gravitational collapse of matter must have at least two LRs, one of which is stable and the other is unstable. Such study showed the great success of topological approach without knowing the specific locations of LRs. Subsequently, such treatment was generalized to a stationary axisymmetric, asymptotically flat black hole \cite{Cunha2020}. The result stated that there is at least one standard unstable LR outside the black hole horizon for each rotation sense. For the static, spherically symmetric black holes in asymptotically flat, dS, and AdS spacetime, such property still held \cite{Wei2020}. Even when more LRs or photon spheres are presented, there is always one more unstable LR or photon sphere. Other relevant studies can also be found in Refs. \cite{Guo2020,Junior2021,Ghosh2021,Junior2021b}.

On the other hand, the equatorial circular orbits for the photons and massive particles are closely related to each other. In Refs. \cite{Delgado2021,Lehebel2022}, it was found that an unstable (stable) LR delimits a region of unstable (stable) TCOs radially above (below) it. Moreover, the corresponding corollary was discussed for both horizonless ultracompact objects and black holes. However one significant difference of TCO from LR is that it not only depends on the black hole parameters, but also on the angular momentum and energy of the particles. It seems that such feature makes it impossible to establish the topology for the TCOs.

However, very recently, it was first noted in our previous work Ref. \cite{Wei2022} that the topology can be well-behaved for the TCOs by constructing an appropriate vector in stationary black holes. Although the angular momentum of the particles modifies the locations of the TCOs, they do not alter the asymptotic behaviors of the constructed vector at the black hole horizon and radial infinity. Thus this suggests that the corresponding topological argument was meaningful and can be applicable to the TCOs. Considering a stationary black hole background, we found that the topological number of TCOs $W=0$ for each fixed angular momentum and is quite different from $W$=-1 of LRs. This suggests that if there exist TCOs, they must appear in pairs for given angular momentum. In particular, stable and unstable TCOs, respectively, have positive and negative winding numbers. For the fixed energy $E$ of the massive particles, the topological number $W=0$ for $0<E<1$ and $W=-1$ for $E>1$, admitting a topological phase transition at $E$=1. These results were exactly confirmed by further applying the topology to the Kerr black holes. For each fixed angular momentum, there may be no TCO, or a pair TCOs as expected. In either case, the topological number keeps zero under this topological configuration of TCOs. Here we refer the representation of the radius and angular momentum of the TCO as the topological configuration.

Nevertheless, the topological argument states that the topological number vanishes for fixed angular momentum. We wonder whether there exist other topological configurations beyond the Kerr black hole case, which possess more than one pair TCOs. By solving this remaining issue, we can further confirm that the topological argument is nontrivial and significant in exploring the TCOs in both GR and modified gravity. In particular, most known black hole solutions in modified gravity are static, spherically symmetric. It is worthwhile extending our previous study of stationary black holes \cite{Wei2022} to these static ones without spin.

Aiming at disclosing different kinds topological configurations and numbers, we in this paper carry out the topological study for the Schwarzschild, scalarized Einstein-Maxwell, and dyonic black holes. For the Schwarzschild black hole, there is at most one pair of TCO. It shares a similar topological configuration as that of the Kerr black hole, whether they have spin or not. While for the scalarized Einstein-Maxwell and dyonic black holes, there could be two pairs TCOs, quite different from the Schwarzschild and Kerr black holes. Furthermore, at small angular momentum, there is no TCO for the scalarized Einstein-Maxwell black hole, and one pair TCOs for the dyonic black hole. So we exhibit three characteristic kinds topological configurations. More surprisingly, all their topological number vanishes regardless of different black hole parameters. This result indicates that such topological argument can be extended to the black hole solutions in modified gravity. Significantly, different topological configurations may correspond to the same topology number, and hence belong to the same topological class.

An outline of the present paper is as follows. In. Sec. \ref{sec2}, we briefly review the topological argument, and apply it to the spherically symmetric, asymptotic flat black holes. The topological number is same as the Kerr black hole for each rotating sense. In Sec. \ref{sec3}, this argument is applied to the Schwarzschild black holes. As expected, the result is consistent with the general analysis. In Sec. \ref{sec4}, the topological study is carried out for the scalarized Einstein-Maxwell black holes. There exhibit five different topological situations of TCOs. For small angular momentum, the TCO does not exist. However for large angular momentum, two pairs TCOs can be observed. Nevertheless, the topological number always keeps zero. In Sec. \ref{sec5}, the dyonic black hole with a quasi-topological term is investigated. Although the quasi-topological term leads to the existence of TCOs for arbitrarily small angular momentum, the total topological charge still vanishes. Finally, we summarize and discuss our results in Sec. \ref{sec6}. In this paper, we adopt the geometrized unit system $c=\hbar=G=1$.

\section{Topological approach}\label{sec2}

In this section, we would like to give a brief introduction of topological approach for the four-dimensional asymptotic flat spacetime, the static and spherically symmetric black holes without spin. The result can also be obtained by adopting a static limit of Ref. \cite{Wei2022}.

First, let us assume the black hole solutions can be described by the following line element
\begin{equation}\label{LE}
    ds^2=g_{tt}dt^2+g_{rr}dr^2+r^2 d\Omega_{2}^{2},
\end{equation}
where $d\Omega_{2}^{2}=d\theta^2+\sin^2\theta d\phi^2$ describes unit 2-sphere. In this black hole background, the Lagrangian of a free test particle reads
\begin{equation}\label{Lag}
    \mathcal{L}=\frac{1}{2}g_{\mu\nu}\dot{x}^\mu \dot{x}^\nu=-\frac{1}{2}\mu^2.
\end{equation}
The dot denotes the derivative with respect to an affine parameter, and $\mu^2=1$, $0$, $-1$ are for the timelike, null, and spacelike geodesics, respectively. Via the Legendre transformation, the Hamiltonian of the test particle can be obtained
\begin{eqnarray}
    \mathcal{H}&=&\pi_\mu \dot{x}^\mu -\mathcal{L}\nonumber\\
    &=&\frac{1}{2}\left(g_{tt}\dot{t}^2+g_{rr}\dot{r}^2+g_{\theta\theta}\dot{\theta}^2+g_{\phi\phi}\dot{\phi}^2\right),
\end{eqnarray}
where $\pi_\mu\equiv\partial\mathcal{L}/ \partial\dot{x}^\mu=g_{\mu\nu}\dot{x}^\nu$ is the corresponding conjugate momentum of the canonical coordinate $x^\mu$.

After a simple rearrangement, the Lagrangian (\ref{Lag}) is reexpressed as
\begin{equation}\label{LR}
    g_{rr}\dot{r}^2+g_{\theta\theta} \dot{\theta}^2+g_{tt}\dot{t}^2+g_{\phi\phi} \dot{\phi}^2+\mu^2=0.
\end{equation}
Since the first two terms are related to the radial and angular motion, we can regard them as the kinetic energy of a test particle. Outside the horizon, both $g_{rr}$ and $g_{\theta\theta}$ are positive, leading to a non-negative kinetic energy as expected. At the same time, the remaining terms on the left-hand-side of equation can be defined as the effective potential
\begin{equation}
    \mathcal{V}=g_{tt}\dot{t}^2+g_{\phi\phi} \dot{\phi}^2+\mu^2.
\end{equation}
It should be emphasized that the motion of particles is completely governed by the effective potential. On the other hand, there are two Killing vectors $\xi^\mu=(\partial_t)^\mu$ and $\psi^\mu=(\partial_\phi)^\mu$ associating with two conservation quantities, the energy $E$ and angular momentum $l$ of the test particle,
\begin{align}
    E & =-g_{\mu\nu}u^\mu\xi^\nu=-g_{tt}\dot{t}, \label{KillE} \\
    l & =g_{\mu\nu}u^\mu \psi^\nu=g_{\phi\phi}\dot{\phi}, \label{KillL}
\end{align}
where $u^\mu$ is the four-velocity of a particle with respect to the affine parameter.

In terms of energy and angular momentum, the effective potential becomes
\begin{equation}\label{Vr}
    \mathcal{V}=\frac{E^2}{g_{tt}}+\frac{l^2}{g_{\phi\phi}}+\mu ^2.
\end{equation}
If \(\mu^2=0\), this expression would precisely reduce to the corresponding effective potential of photon. Another feature is that the effective potential is symmetric under $l\to-l$. Without loss of generality, we only focus on the positive angular momentum $l$. On the other hand, the formula (\ref{Vr}) is a quadratic form of energy \(E\), and can be factorized as \cite{Wei2022}
\begin{equation}\label{Vf}
    \mathcal{V}=\frac{1}{g_{tt}}(E-e_1)(E-e_2),
\end{equation}
with
\begin{equation}\label{e1e2}
    e_1=\sqrt{\frac{-g_{tt} \left(l^2+g_{\phi\phi}  \mu^2\right)}{g_{\phi\phi}}},\quad e_2=-\sqrt{\frac{-g_{tt} \left(l^2+g_{\phi\phi} \mu^2\right)}{g_{\phi\phi}}}.
\end{equation}
The timelike circular orbit of a massive particle ($\mu^2=1$) requires
\begin{equation}\label{tco}
    \mathcal{V}(r)=0,  \quad  \frac{\partial \mathcal V(r)}{\partial r}=0.
\end{equation}
For given $l$, one can obtain the radius of the TCOs through the second condition. Then the energy of the particle relating with the TCOs will be determined by the first condition. Considering that $e_2$ is negative, we abandon it here. As a result, the conditions (\ref{tco}) turn to
\begin{equation}\label{contco}
    E=e_1, \quad  \partial_r\, e_1=0.
\end{equation}
Compared with LR, the TCO not only depends on the black hole parameter, but also on the energy and angular momentum of the particle. Solving above conditions, one obtains the energy and angular momentum of TCOs on the equatorial plane \cite{Witek}
\begin{equation}\label{ltco}
    l_{t}=\sqrt{\frac{r^3 g_{tt}'(r)}{2\,g_{tt}(r)-r\,g_{tt}'(r)}},\quad E_{t}= \frac{-\sqrt{2}\,g_{tt}(r)}{\sqrt{ r\, g_{tt}'(r)-2\, g_{tt}(r)} }.
\end{equation}
As previously stated, the TCOs are completely controlled by $e_1$. Moreover, we can construct a vector that is related to the topology of TCOs. Following Ref. \cite{Cunha2020}, it is convenient to introduce the following vector
\begin{equation}\label{PhiR}
    \phi^r=\frac{\partial_r e_1}{\sqrt{g_{rr}}},\quad \phi^\theta=\frac{\partial_\theta e_1}{\sqrt{g_{\theta\theta}}}.
\end{equation}
If the considered spacetime has $\mathcal{Z}_2$ symmetry in $\theta$, the zero points of vector $\phi$ will locate at $\theta=\pi/2$ and $\partial_r e_1=0$, which exactly correspond to the equatorial TCOs.

On the other hand, following Duan's $\phi$-mapping topological current theory \cite{Duan1984}, a point-like particle corresponding to the zero point of a vector filed $\phi$ can be endowed with a topological charge. The conservation of particle number is well guaranteed by the total topological charge. As shown above, the TCOs exactly locate at the zero points of vector $\phi$. Thus, we can endow each TCO with a topological charge. This allows us to establish the topology for the TCOs. Then, the topological properties will be uncovered as expected. Following Ref. \cite{Duan1984}, the topological current associated with the topological charge reads
\begin{equation}
    j^\mu=\frac{1}{2\pi}\epsilon^{\mu\nu\rho} \epsilon_{ab}\frac{\partial n^a}{\partial x^\nu}\frac{\partial n^b}{\partial x^\rho},
\end{equation}
where $x^\mu=(t,r,\theta)$ and $n^a=( \frac{\phi^r}{|\phi|}, \frac{\phi^\theta}{|\phi|})$ is the unit vector of \(\phi\). It is easy to check that this current is conserved, i.e., $\partial_{\mu}j^{\mu}$=0. After a simple algebra, one reaches
\begin{equation}
    j^\mu=\delta^2(\phi)J^{\mu}\left(\frac{\phi}{x}\right),
\end{equation}
with Jacobi tensor $\epsilon^{ab}J^{\mu}\left(\frac{\phi}{x}\right)=\epsilon^{\mu\nu\rho}\partial_{\nu}\phi^a\partial_{\rho}\phi^b$. Significantly, $j^{\mu}$ is nonzero only at the zero points of the vector $\phi$. Denoting the $i$-th zero point as $\vec{x}=\vec{z}_{i}$, we have the density $j^0$ of the topological current \cite{Duan1984}
\begin{equation}
    j^0=\sum_{i}^{N}=\beta_i\eta_i\delta^2(\vec{x}-\vec{z}_{i}),
\end{equation}
where $\beta_i$ and $\eta_i$ are the Hopf index and Brouwer degree of the $i$-th zero point. By integrating the density $j^0$ of the topological current over the giving region $\Sigma$, one obtains the topological number
\begin{equation}
    W=\int_\Sigma j^0 d^2 x=\sum_{i}^{N}\beta_i\eta_i=\sum_{i}^{N}w_i.
\end{equation}
Here $w_i$ denotes the winding number of the $i$-th zero point of the vector $\phi$ enclosed in $\Sigma$. The topological number can be calculated by counting the deflection angle $\Omega$ of the vector direction along the counterclockwise closed path $\mathcal{I}=\partial\Sigma$
\begin{align}
    W=\frac{1}{2\pi}\oint_\mathcal{I} \mathrm{d}\Omega.
\end{align}
From a local perspective, each zero point is characterized by a winding number, and employing which these zero points can be classified into several classes with different topological properties. Conversely, if the considered region $\Sigma$ covers all the possible parameter space, the global topological property will be uncovered by the number $W$.

We examine the topological properties of the TCOs for the static and spherically symmetric black holes, and find that there is no sudden change when the black hole spin tends to vanish. For convenience, we summarize the main results as belows: i) The stable and unstable TCOs have winding number $w$=+1 and -1, respectively.  ii) The outermost TCO with $w$=1 is stable, while the innermost one has $w$=-1 and is unstable. iii) The marginally stable circular orbit (MSCO) acts as bifurcation point \cite{Fu2000}, and has vanished winding number. iv) The topological number $W=0$, implies that the TCOs always come in pairs if they exist.

Although it seems that the results keep the same as that of the rotating black holes, we will show that there exist distinct topological configurations of the TCOs for different nonrotating black holes. More than one pairs TCOs present and the ISCOs will be not necessarily the bifurcation points. These shall greatly enlarge our understanding on the topological properties of the TCOs. For this purpose, in the following sections, we would like to examine the topology for three characteristic black hole solutions.

\section{Schwarzschild black holes}\label{sec3}

In this section, we shall carry out the topological study of the TCOs for the Schwarzschild black hole.

The Schwarzschild black hole is a static spherically symmetric vacuum solution of the Einstein field equation, and it can be described by the line element (\ref{LE}) with
\begin{equation}
    g_{tt}=-\left(1-\frac{2M}{r}\right), \quad g_{rr}=\left(1-\frac{2M}{r}\right)^{-1},
\end{equation}
where $M$ is the black hole mass and the corresponding event horizon locates at $r_h=2M$.

From (\ref{Vr}), the effective potential $\mathcal{V}(r)$ reads
\begin{equation}
    \mathcal{V}(r)= \frac{l^2 \csc^2\theta}{r^2}+\frac{r E^2}{2M-r}+\mu ^2.
\end{equation}
Reformulating it, one gets $e_1$ and $e_2$ via the equation (\ref{e1e2})
\begin{equation}
    e_{1,2}=\pm\sqrt{\frac{(r-2M)(r^2\mu^2+l^2\csc^2\theta)}{r^3}}.
\end{equation}
Following the definition (\ref{PhiR}), the vector \(\phi\) is
\begin{equation}
    \phi^r=\frac{Mr^2+(3M-r)l^2\csc^2\theta}{r^3\sqrt{r^2+l^2\csc^2\theta}}, \quad
    \phi^\theta=-\frac{l^2\cot\theta \csc^2\theta\sqrt{r-2M}}{r^{5/2}\sqrt{r^2+l^2\csc \theta}},
\end{equation}
where we have taken \(\mu^2=1\) for the timelike geodesics.

Solving the zero points of the vector, namely $\phi^r=\phi^\theta=0$, we obtain the angular momentum of the TCOs
\begin{equation}
    l_t= \sqrt{\frac{r^2 M}{r-3 M}},
\end{equation}
for the Schwarzschild black holes. Further solving $\partial_{r}l_{t}=0$, one obtains the radius and angular momentum of the ISCO or MSCO
\begin{equation}
    r_{ISCO}=6M \quad \text{and}\quad  l_{ISCO}=2\sqrt{3}M.
\end{equation}
Note that for the Schwarzschild black hole, the ISCO and MSCO coincide, and thus we will not distinguish them here. In what follows, we would like to figure out three characteristic cases according to the angular momentum:
\begin{itemize}
    \item $0\leq l<l_{ISCO}$,
    \item \(l=l_{ISCO}\),
    \item \(l_{ISCO}<l<\infty\).
\end{itemize}
For simplicity, we shall take $M=1$ for our following study.

\subsection{Topology of TCOs and winding number}

\begin{figure}[h]
    \centering
    \subfigure[]{\includegraphics[width=6cm]{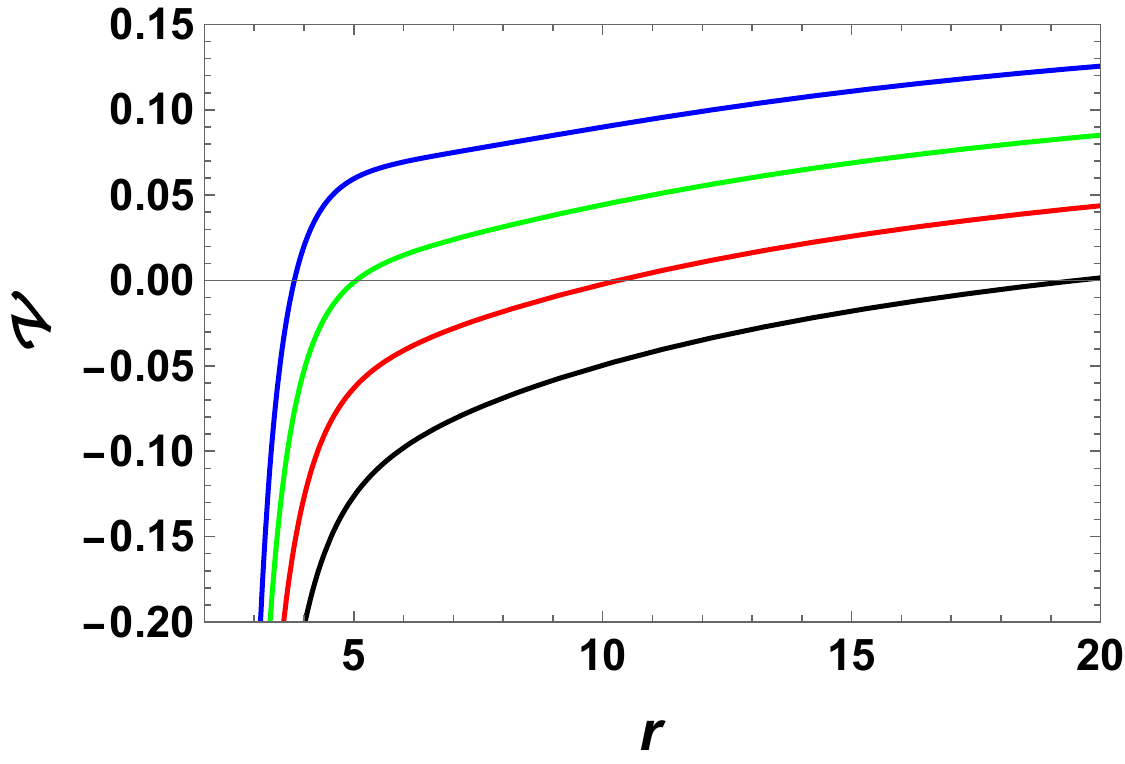}\label{schC1V}}
    \subfigure[]{\includegraphics[width=6cm]{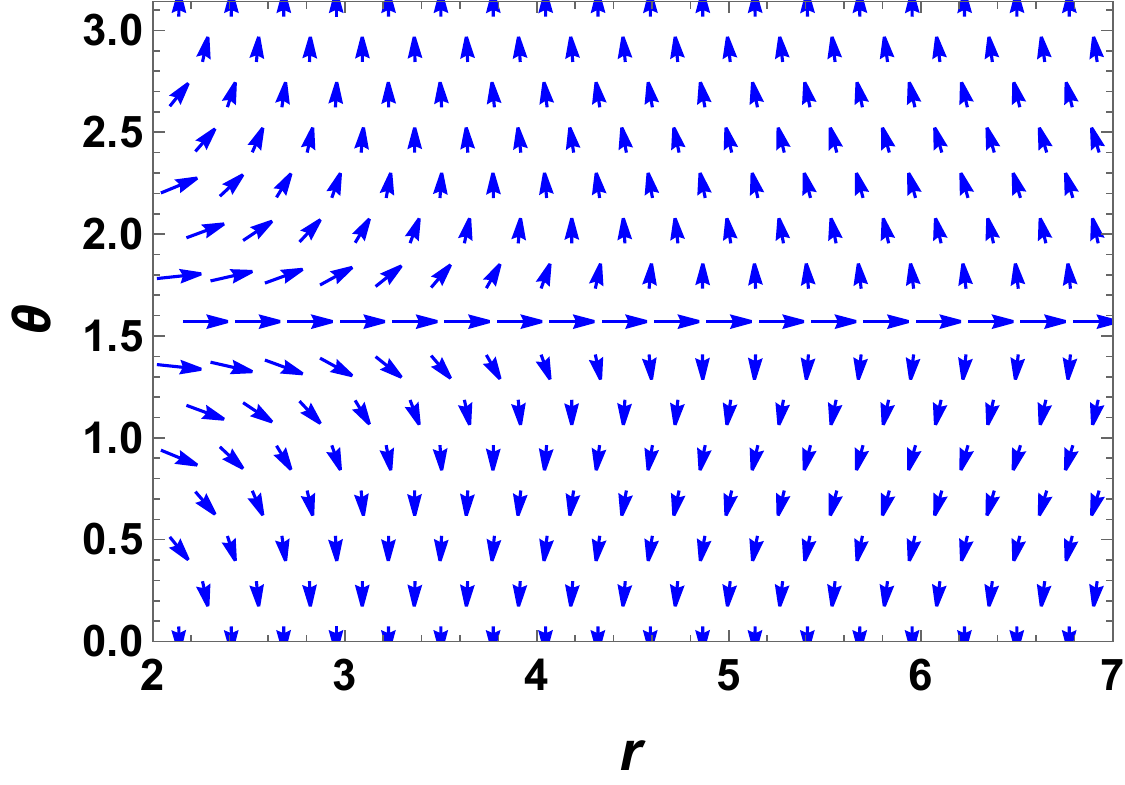}\label{schC1N}}
    \caption{The effective potential \(\mathcal{V}(r)\) and unit vector field of $\phi$ for the Schwarzschild black holes for the first case with \(l=3.2\). (a) \(\mathcal{V}(r)\) as a function of $r$ with $E=0.9$, $0.92$, $0.94$, and $0.96$ from bottom to top. (b) The unit vector field $n$ on a portion of the $\theta$-$r$ plane.}\label{schCase1}
\end{figure}

First, let us consider $0<l=3.2<l_{ISCO}$. The effective potential \(\mathcal{V}(r)\) is plotted in Fig. \ref{schC1V} when $E=0.9$, $0.92$, $0.94$, and $0.96$ from bottom to top, respectively. For each curve, we find that, \(\mathcal{V}(r)\) increases monotonically with $r$. So for this case, there is no TCO. Also, we show the unit vector field $n$ on a portion of the $\theta$-$r$ plane in Fig. \ref{schC1N}. Obviously, the vector is outwards at $\theta=0$ and $\pi$. On the equatorial plane \(\theta=\pi/2\), the direction of the vector is always toward to the right, and no zero point can be found. So the topological number must vanish, i.e., $W=0$.

\begin{figure}[h]
    \centering
    \subfigure[]{\includegraphics[width=6cm]{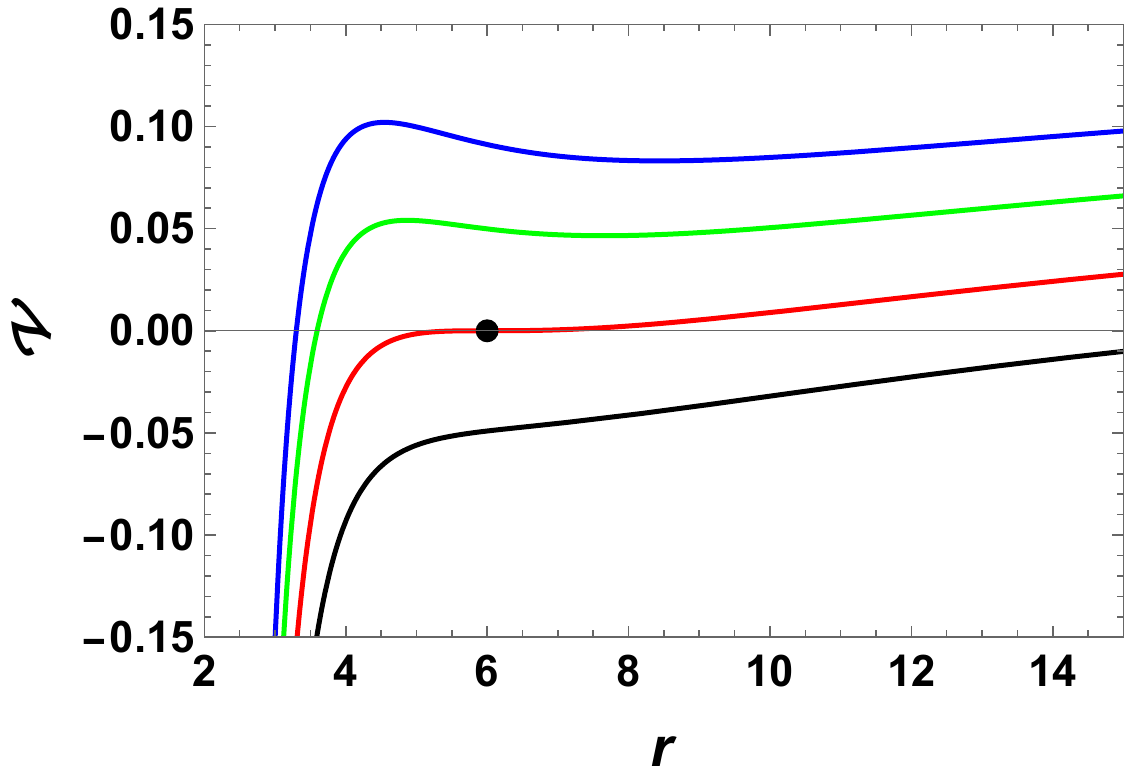} \label{schC2V}}
    \subfigure[]{\includegraphics[width=6cm]{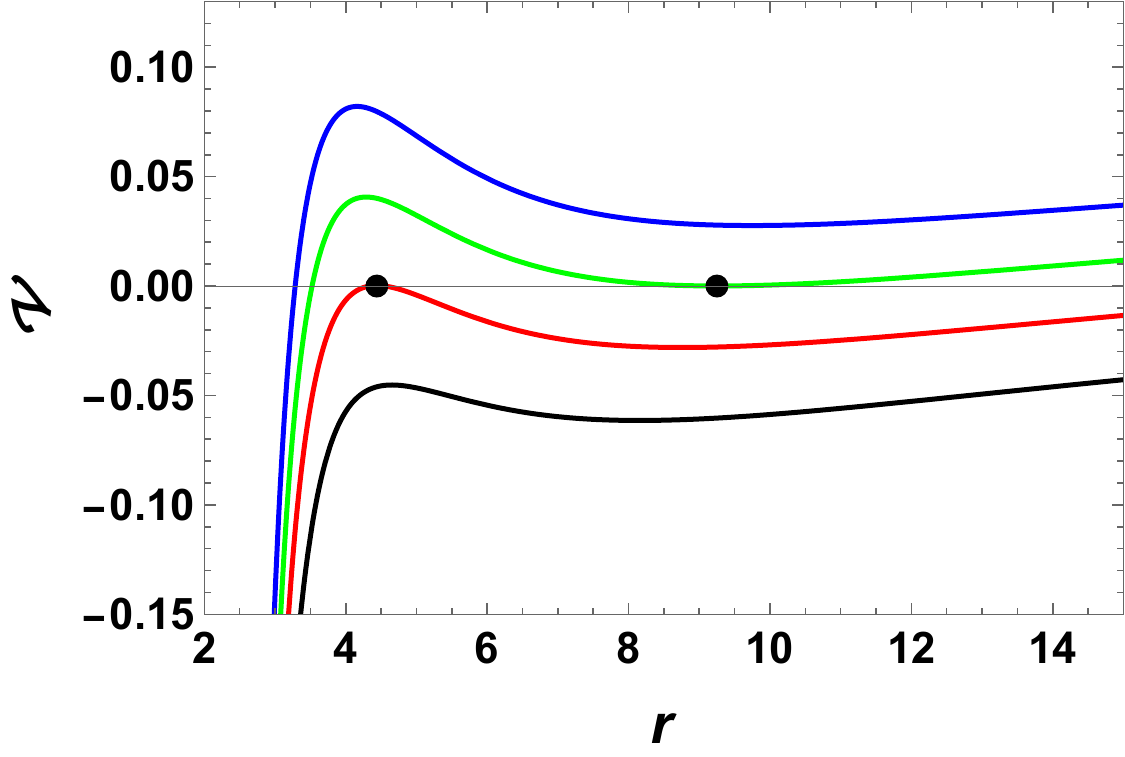}\label{schC3V}}\\
    \subfigure[]{\includegraphics[width=6cm]{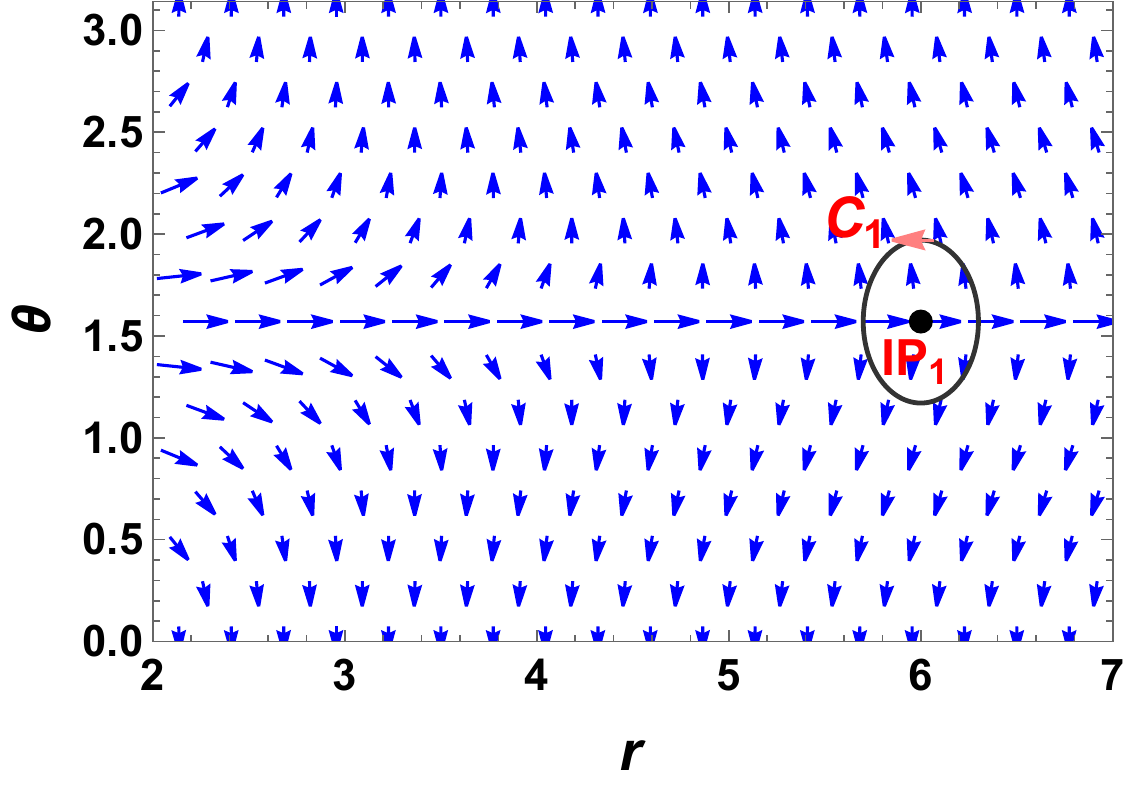}\label{schC2N}}
    \subfigure[]{\includegraphics[width=6cm]{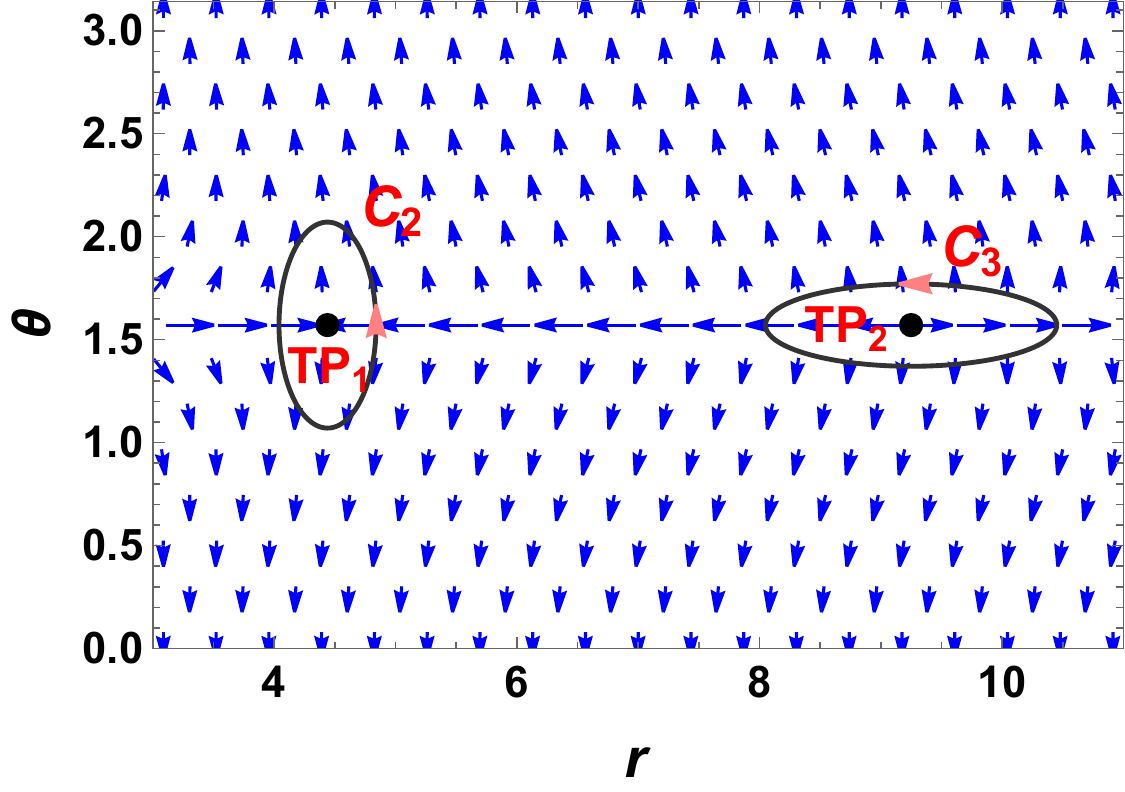}\label{schC3N}}\\
    \subfigure[]{\includegraphics[width=6cm]{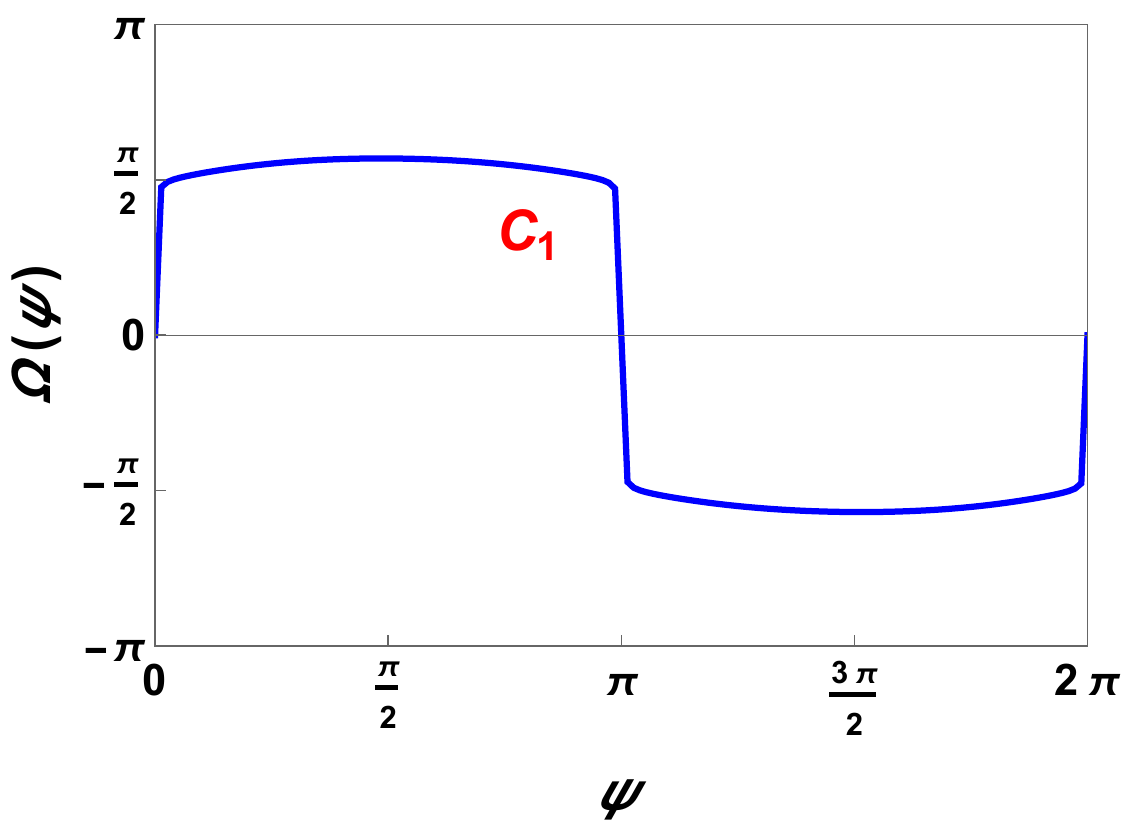}\label{schC2A}}
    \subfigure[]{\includegraphics[width=6cm]{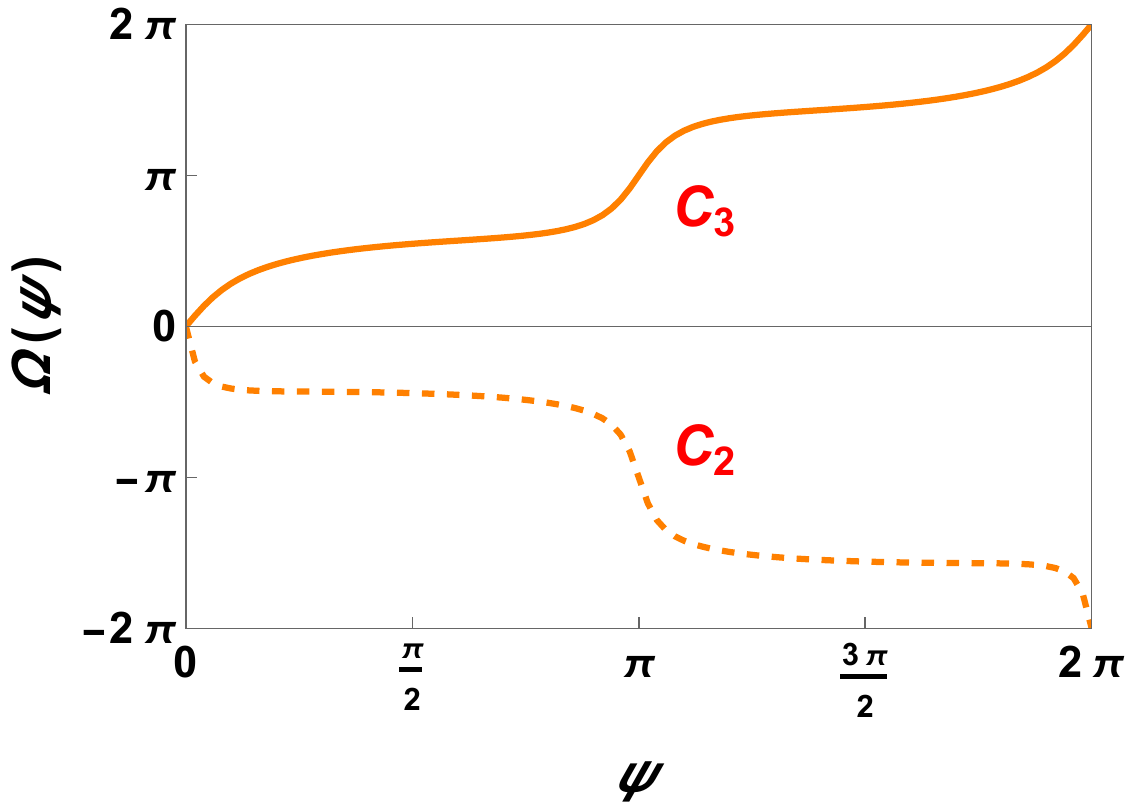}\label{schC3A}}
    \caption{The effective potential $ \mathcal{V} $, unit vector field $n$  and deflection angle $ \Omega(\psi) $ for the Schwarzschild black holes. (a) Effective potential $ \mathcal{V} $ for the second case with $l=l_{ISCO}$. The energy $E=0.91$, $0.925$, $0.9428$ and $0.96$ from bottom to top. (b) Effective potential $\mathcal{V}$ for the third case with $l=3.7$. The energy $E=0.942$, $0.9535$, $0.9649$ and $0.978$ from bottom to top. (c) The unit vector field $n$ on a portion of the $\theta$-$r$ plane with $l=l_{ISCO}$. ``$IP_1$" denotes the ISCO at $r_{ISCO}$=6. The closed loop $C_1$ has parametric coefficients ($c_0$, $c_1$, $c_2$)=($r_{ISCO}$, 0.4, 0.5). (d) The unit vector field $n$ on a portion of the $\theta$-$r$ plane with $l=3.7$. ``$TP_1$" and ``$TP_2$" are two TCOs located at $r_t$=4.44 and 9.25. The closed loops $C_2$ and $C_3$ have parametric coefficients ($c_0$, $c_1$, $c_2$)=(4.44, 0.4, 0.5) and (9.25, 1.2, 0.2). (e) Deflection angle \(\Omega(\psi)\) along $C_1$. (f) Deflection angle $\Omega(\psi)$ along $C_2$ and $C_3$.}
\end{figure}

For the second characteristic case, we take  \(l=l_{ISCO}=3.4641\). The effective potential and the unit vector field $n$ on a portion of the $\theta$-$r$ plane are shown in Figs. \ref{schC2V} and \ref{schC2N}. The ISCO exactly locates at $r=6$ with $E=2\sqrt{2}/3$. From Fig. \ref{schC2V}, it can be found that for $E>2\sqrt{2}/3$, no extremal point is present for the effective potential \(\mathcal{V}(r)\), while two extremal points are present for $ E<2\sqrt{2}/3$. In Fig. \ref{schC2N}, it is easy to find that the unit vector field $n$ has a similar behavior as that given in Fig. \ref{schC1N}. Although the direction of $n$ seems to be always towards to the right on the equatorial plane, it indeed vanishes at the point marked with the dot, and $\partial_{r,r}\mathcal{V}=0$ is satisfied. Here we wonder whether the winding number still vanishes as that case of $l=3.2$. In order to answer this question, we turn to evaluate its winding number  by constructing a closed loop $C_1$ with following parameterized form \cite{Wei2021}
\begin{equation}\label{paraeq}
    \begin{cases}
    r=c_1 \cos\psi +c_0, \\
    \theta=c_2\sin\psi+\frac{\pi}{2},
    \end{cases}
\end{equation}
where ($c_0$, $c_1$, $c_2$)=(3.4641, 0.4, 0.5). Note that all the constructed closed loops below will be parameterized with this form while with different values of ($c_0$, $c_1$, $c_2$). Along the closed loop, the deflection angle $\Omega(\psi)$ can be calculated by
\begin{equation}
    \Omega(\psi)=\int_C \epsilon_{ab}\,n^a\, \mathrm{d}n^b.
\end{equation}
The winding number shall be $w=\Omega(2\pi)/2\pi$ as expected. We list the deflection angle $\Omega(\psi)$ in Fig. \ref{schC2A}. With the increase of $\psi$ from 0 to $2\pi$, we see that $\Omega(\psi)$ first increases, then decreases, and finally increases. Nevertheless, $\Omega(2\pi)$ vanishes, strongly implying that the winding number $w=0$ for the ISCO.

Next we turn our attention to $l=3.7>l_{ISCO}$. The effective potential $\mathcal{V}$ is plotted in Fig. \ref{schC3V}. For different values of energy, two extremal points can be observed on each curve. However, they do not denote the TCOs unless they have vanished potential. According to it, we find that there are two TCOs marked with dots at $r$=4.44 and 9.25, which correspond to $E=0.9535$ and $0.9649$, respectively. Also, from the behaviors of the effective potential, one easily reaches that the TCO at $r=4.44$ is local unstable, while the other one is stable. As we shall see, this result will also be confirmed by their winding numbers.

The unit vector field $n$ is also described on a portion of the $\theta$-$r$ plane in Fig. \ref{schC3N}. Obviously, there are two zero points, which are exactly consistent with these shown in Fig. \ref{schC3V}. In order to calculate their winding numbers, we construct two closed loops $C_2$ and $C_3$ parametrized by the form (\ref{paraeq}) with ($c_0$, $c_1$, $c_2$)=(4.44, 0.4, 0.5) and (9.25, 1.2, 0.2). Then we show $\Omega(\psi)$ for them in Fig. \ref{schC3A}. With the increase of $\psi$, $\Omega(\psi)$ increases along $C_3$, while decreases $C_2$. The winding number is easily got, for example $w$=1 for $TP_2$ and -1 for $TP_1$, which implies that a positive or negative winding number corresponds to a local stable or unstable TCO as expected. As a result, the total topological number $W$=+1-1=0, keeping the same with the first two cases.

\subsection{Topological configuration}

\begin{figure}[h]
    \centering
    \subfigure[]{\includegraphics[width=6cm]{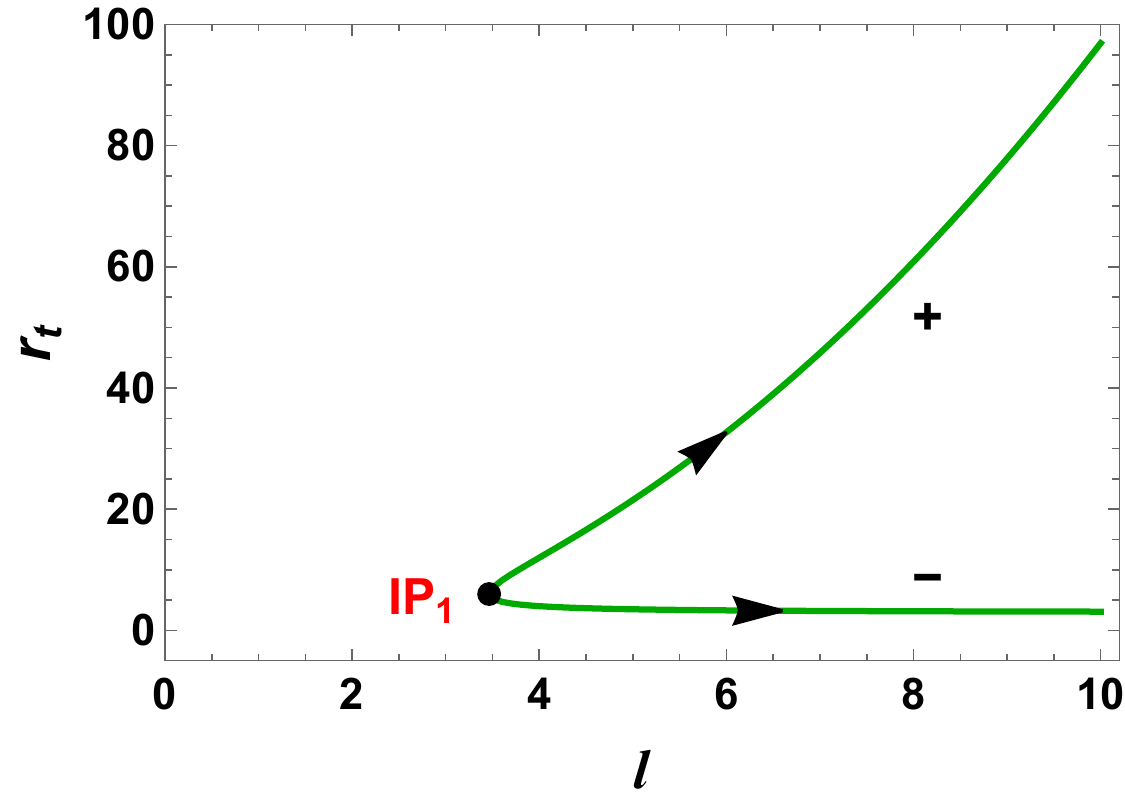}\label{schRL}}
    \subfigure[]{\includegraphics[width=6cm]{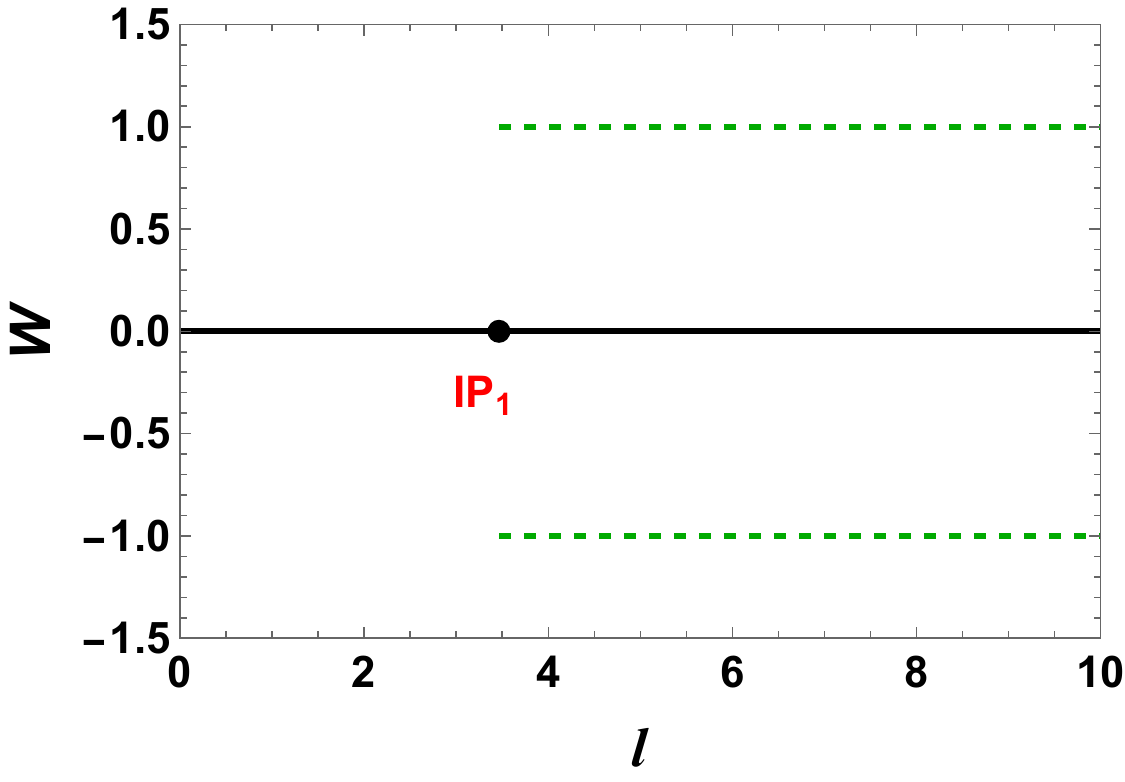}\label{schWL}}
    \caption{(a) The evolution of the TCO radius \(r_t\) as a function of angular momentum \(l\) for the Schwarzschild black holes. Obviously, the point ${\rm IP}_1$ is a generated point. The signs $\pm$ denote the winding number are $\pm1$ for these TCO branches. (b) The topological number (solid black line) and winding number of TCO branches (dashed green lines). For different values of $l$, we see the topological number $W$ always vanishes.}
\end{figure}

After studying the topological charge of ISCO and TCO, we concentrate on the evolution of TCO radius \(r_t(l)\), from the idea that the angular momentum can be treated as a time control parameter \cite{Wei2022}. Expanding the angular momentum $l_t$ of the zero points of the vector $\phi$ at $r_{ISCO}$, one has
\begin{align}
    l_t & =2\sqrt{3}+\frac{1}{12\sqrt{3}} (r-6)^2+\mathcal{O}\left((r-6)^3\right).
\end{align}
Since $l_t''(r_{ISCO})=1/(12\sqrt{3})>0$, the bifurcation point must be a generated point. In order to make it clearer, we display the radius $r_t$ of the TCOs as a function of $l$ in Fig. \ref{schRL}. For small $l$, no branch of the TCO can be found. While after the ISCO point, two TCO branches emerge from the ISCO with opposite winding numbers. Such behavior obviously addresses that the ISCO is a generated point with the angular momentum. More interestingly, whether there are TCOs or not, the total topological number always vanishes for arbitrary values of the angular momentum. Furthermore, we sketch the behavior of the winding number and topological number in Fig. \ref{schWL}.

In summary, we in this section observe that there is no, or one pair TCOs at small and large angular momenta. Such topological configuration is similar to that of the Kerr black hole and thus they share the same topological number, which further implies that this topological argument is independent of the black hole spin. Moreover, these results exactly support our general conclusion given in Sec. \ref{sec2}.

\section{Scalarized Einstein-Maxwell black holes}\label{sec4}

In this section, we would like to consider another characteristic example, the scalarized Einstein-Maxwell black holes. Even though their total topological number is the same as the Schwarzschild black holes, the topological configurations of the TCOs are quite different.

\subsection{Scalarized Einstein-Maxwell black holes}

The Einstein-Maxwell-scalar model describes a real scalar field \(\phi\) coupling to Einstein's gravity and Maxwell's electromagnetism. The scalarized Einstein-Maxwell black hole can be described by the following action \cite{Silva2017,Konoplya2019}
\begin{equation}\label{actionEMS}
    \mathcal{S}=\int d^4x \sqrt{-g}(R-2g^{\mu\nu}\partial_\mu \phi \partial_\nu\phi-f(\phi)F_{\mu\nu}F^{\mu\nu}),
\end{equation}
where \(R\), \(\phi\), and \(F_{\mu\nu}=\partial_\mu A_\nu -\partial_\nu A_\mu\) are the Ricci scalar and scalar field, and Maxwell tensor. The last term represents a non-minimal coupling term between scalar field and the Maxwell electric field.

\begin{figure}[h]
    \centering
    \includegraphics[width=8cm]{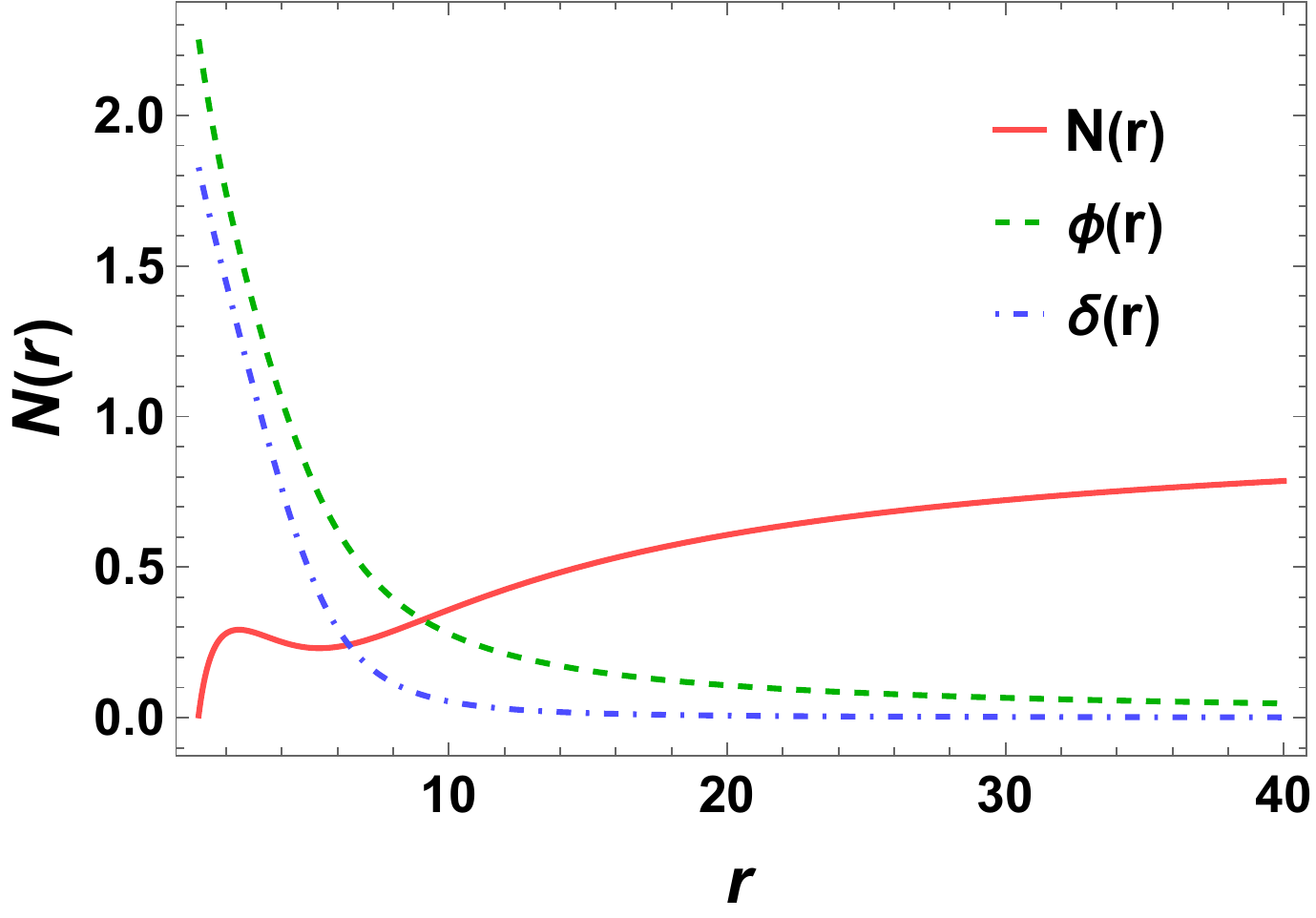}
    \caption{The numerical results of functions \(N(r)\), \(\phi(r)\), and \(\delta(r)\) for scalarized Einstein-Maxwell black holes.}
    \label{EMSfun}
\end{figure}

The line element of the spherically symmetric scalarized Einstein-Maxwell black holes is assumed to be
\begin{equation}
    ds^2=-N(r) e^{-2\delta(r)}dt^2+\frac{dr^2}{N(r)}+r^2 d\Omega^2,
\end{equation}
where the metric function $N(r)$ and $\delta(r)$ are only radially dependent. The four-potential $A_\mu$ of the electromagnetic field is
\begin{equation}
    A_\mu(x) dx^\mu=V(r)dr.
\end{equation}
The effective Lagrangian in the Einstein-Maxwell-scalar model are \cite{Fernandes2019}
\begin{equation}\label{Leff}
    \mathcal{L}_{eff}=-\frac{1}{2} e^{-\delta} \left(r N'+N-1\right) -\frac{1}{2}e^{-\delta}r^2N\phi'(r)^2+\frac{1}{2}e^\delta f(\phi)r^2V'(r)^2.
\end{equation}
Here the radially dependent is omitted for notation simplicity. By making use of the Lagrangian, the equations of motion read \cite{Fernandes2019}
\begin{align}
    & N'-\frac{1-N}{r}=-\frac{Q^2}{r^3 f(\phi)}- r(\phi')^2 N, \;\quad \delta'=-r(\phi')^2, \label{diffeq1} \\
    & (r^2N\phi')'=-\frac{f'(\phi)Q^2 }{2f^2(\phi)r^2}-r^3(\phi')^3N,  \quad V'=\frac{Q}{f(\phi)r^2}e^{-\delta}.\label{diffeq2}
\end{align}
Further, we assume this black hole spacetime is asymptotic flat
\begin{equation}
    \lim_{r\to\infty} N(r)=1,\quad \lim_{r\to\infty}\delta(r)=0.
\end{equation}
To numerically solve the differential Eqs. (\ref{diffeq1}) and (\ref{diffeq2}), the exponential coupling are chosen as \cite{Guo2021b}
\begin{equation}
    f(\phi)=e^{\alpha \phi^2},
\end{equation}
with $\alpha=0.9$. The value of scalar field at the event horizon
\begin{equation}
    \phi({r_h})=2.25859.
\end{equation}
The numerical results of the metric functions \(N(r)\), \(\delta(r)\), and \(\phi(r)\) are exhibited in Fig. \ref{EMSfun} by taking $r_h=1$. We observe that \(\delta(r)\) and \(\phi(r)\) decrease with $r$, while \(N(r)\) shows a nonmonotonic behavior. After obtaining these functions, the explicit form of \(\mathcal{V}(r)\) and energy \(e_1\), as well as the vector $\phi$ will be numerically given. Other studies concerning the black hole solutions and potential observations of multi-photon sphere can be found in Refs. \cite{Guo2021b,Fernandes,Pappas,Gan2021,Guo2022ba}.

\subsection{Topology of TCOs and winding number}

Now, let us turn to examine the topology of the TCOs for the scalarized Einstein-Maxwell black holes.

As we have shown in the last section, ISCOs or MSCOs acting as bifurcation points, have an important impact on the topological configurations of TCOs. So it is key to determine them first. By solving the conditions \(\mathcal{V}(r)=\mathcal{V}'(r)=0\)=\(\mathcal{V}''(r)=0\), or alternatively,
\begin{equation}\label{iscophi}
    \phi^{r}(r)=0 \quad  \text{and} \quad  \phi^{r\prime}(r)=0,
\end{equation}
we obtain the locations of ISCO and MSCO, which are given by
\begin{align}
    r_{ISCO} & =2.3294,  \quad l_{ISCO}=6.4043,   \nonumber \\
    r_{MSCO} & =15.753,   \quad l_{MSCO}=12.314.
\end{align}
Further, they must satisfy
\begin{equation}\label{eqV3}
    \mathcal{V}''(r)=0, \quad  \mathcal{V}'''(r)>0,
\end{equation}
or,
\begin{equation}\label{SCOMCO}
    \phi^{r\prime}(r)=0, \quad \phi^{r\prime\prime}(r)>0.
\end{equation}
A simple algebra gives
\begin{equation}
    \mathcal{V}'''(r)=\frac{E-e_2(r)}{-g_{tt}} \sqrt{g_{rr}}\, \phi_r''(r),
\end{equation}
which implies that $\mathcal{V}'''(r) \sim\phi_r''(r)$. Note that the ISCO also satisfies condition (\ref{eqV3}), but it has the smallest radius among all stable TCOs. Besides, the significant discrepancy between the ISCO and MSCO is that the MSCO can be continuously connected to spatial infinity by a set of stable TCOs, whereas the ISCO fails.

\begin{figure}[h]
    \centering
    \subfigure[]{\includegraphics[width=6cm]{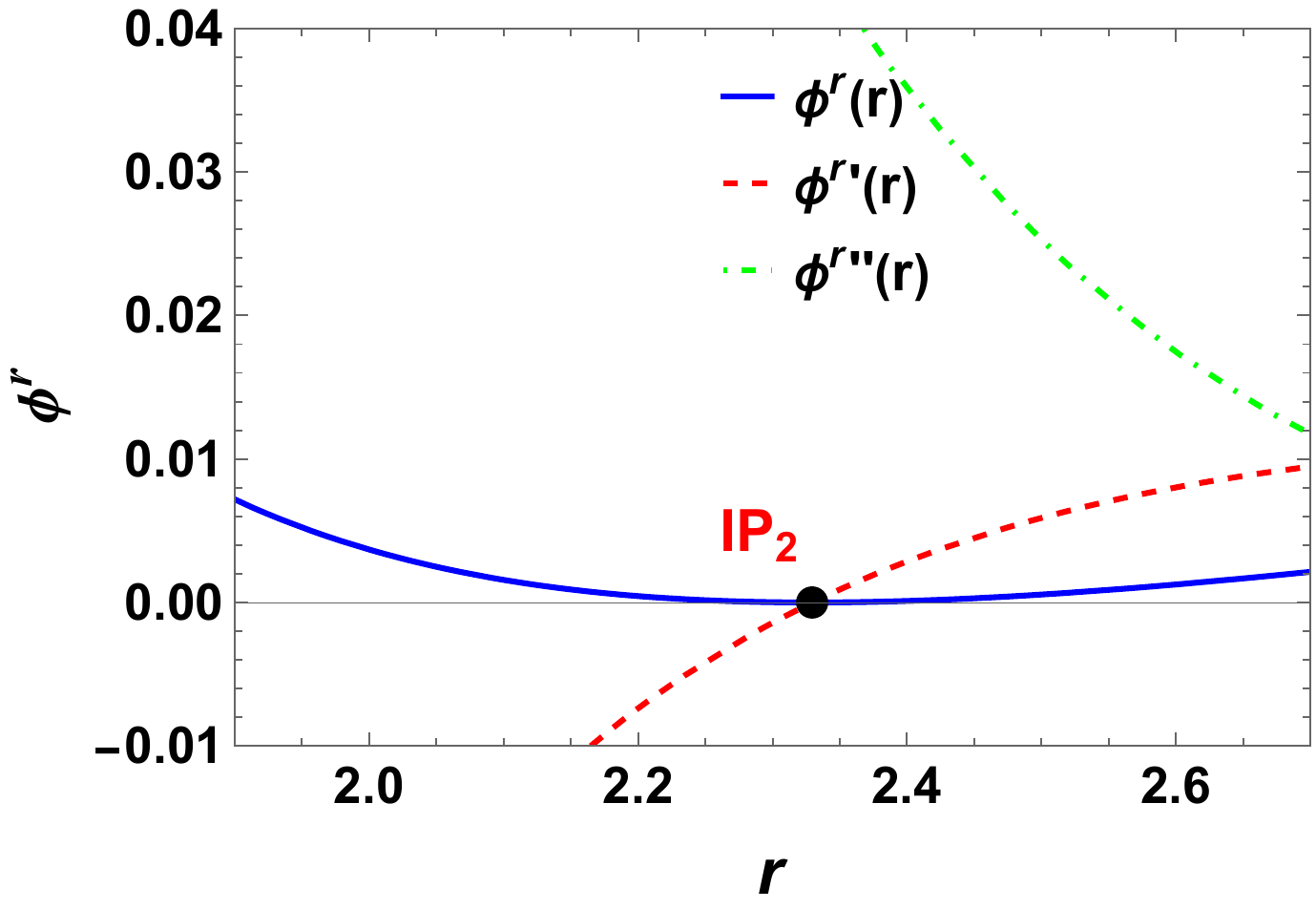}}
    \subfigure[]{\includegraphics[width=6cm]{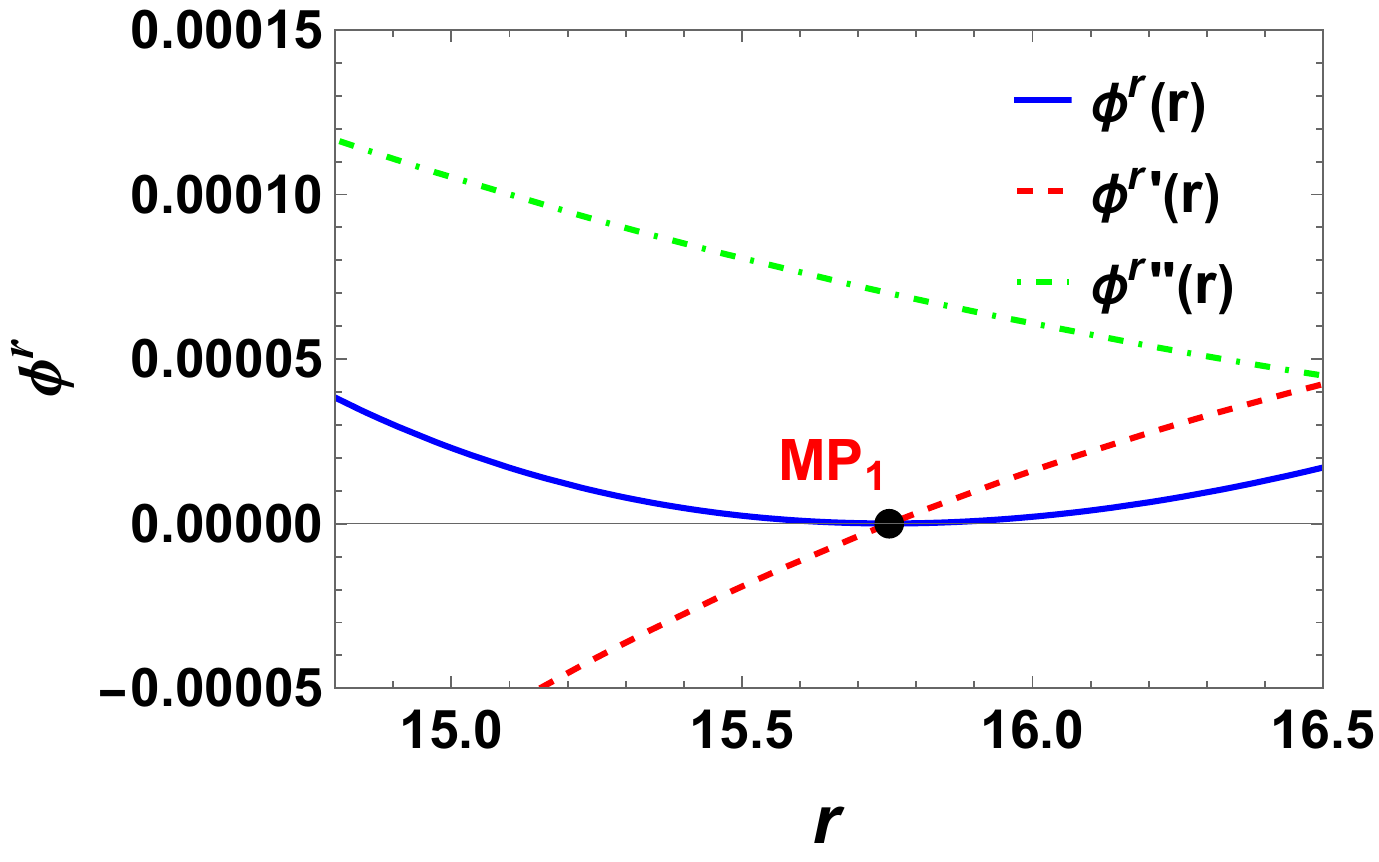}}
    \caption{The function $\phi^r(r)$ and its derivatives $\phi^{r\prime}(r)$ and $\phi^{r\prime\prime}(r)$ for scalarized Einstein-Maxwell black holes. (a) $l=l_{ISCO}$. (b) $l=l_{MSCO}$. The black dots ${\rm IP}_2$ and ${\rm MP}_1$ stand for the locations of the ISCO or MSCO, respectively.}
    \label{EMSphi}
\end{figure}

To check whether the ISCO and MSCO hold the condition (\ref{SCOMCO}), we calculate the derivatives of \(\phi^r(r)\) at the \(r_{ISCO}\) and \(r_{MSCO}\)
\begin{equation}\label{resofphi}
    \begin{split}
    \phi^{r\prime}(r_{ISCO})&=0, \quad \phi^{r\prime \prime}(r_{ISCO})=0.0455786, \\
    \phi^{r\prime}(r_{MSCO})&=0, \quad \phi^{r\prime\prime}(r_{MSCO})=0.00007001.
    \end{split}
\end{equation}
To show this result more clearly, we plot \(\phi^r(r)\) and its derivatives in Fig. \ref{EMSphi}.

According to the values of $l_{ISCO}$ and $l_{MSCO}$, we divide the parameter range of the angular momentum into the following five types:
\begin{itemize}
    \item $0\leq l<l_{ISCO}$,
    \item $l=l_{ISCO}$,
    \item $l_{ISCO}<l<l_{MSCO}$,
    \item $l=l_{MSCO}$,
    \item $l_{MSCO}<l$.
\end{itemize}

\begin{figure}[htb]
    \centering
    \subfigure[]{\includegraphics[width=6cm]{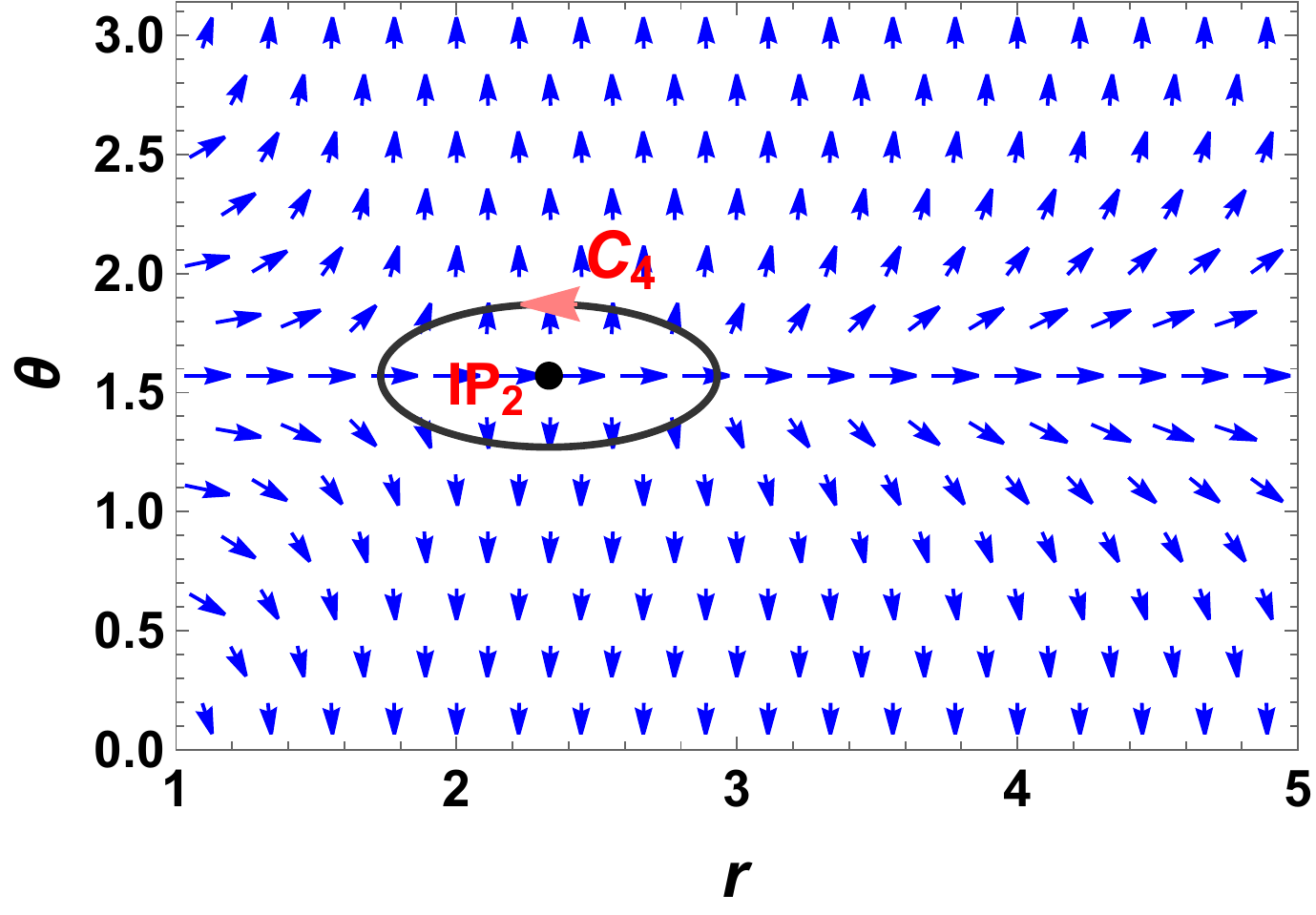}\label{emsC2N}}
    \subfigure[]{\includegraphics[width=6cm]{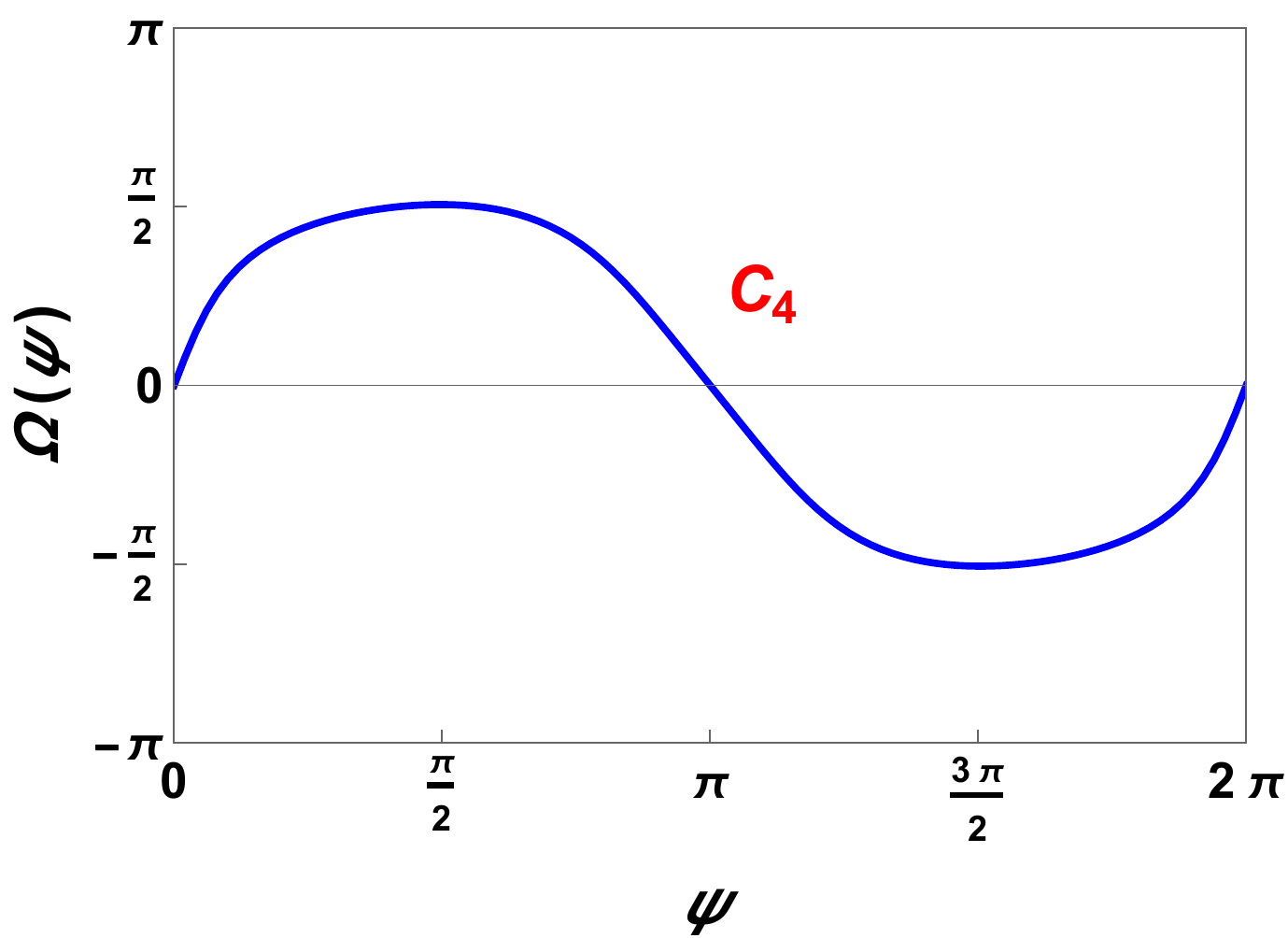}\label{emsC2A}}
    \caption{The second case with $ l=l_{ISCO} $ for the scalarized Einstein-Maxwell black holes. (a) The unit vector field $n$ on a portion of the $\theta$-$r$ plane. ``$IP_2$" denotes the ISCO of the black hole with $r$=2.329. The closed loops $C_4$ has parametric coefficients ($c_0$, $c_1$, $c_2$)=(2.329, 0.6, 0.3). (b) Deflection angle \(\Omega(\psi)\) along $C_4$.}
\end{figure}

\begin{figure}[htb]
    \centering
    \subfigure[]{\includegraphics[width=6cm]{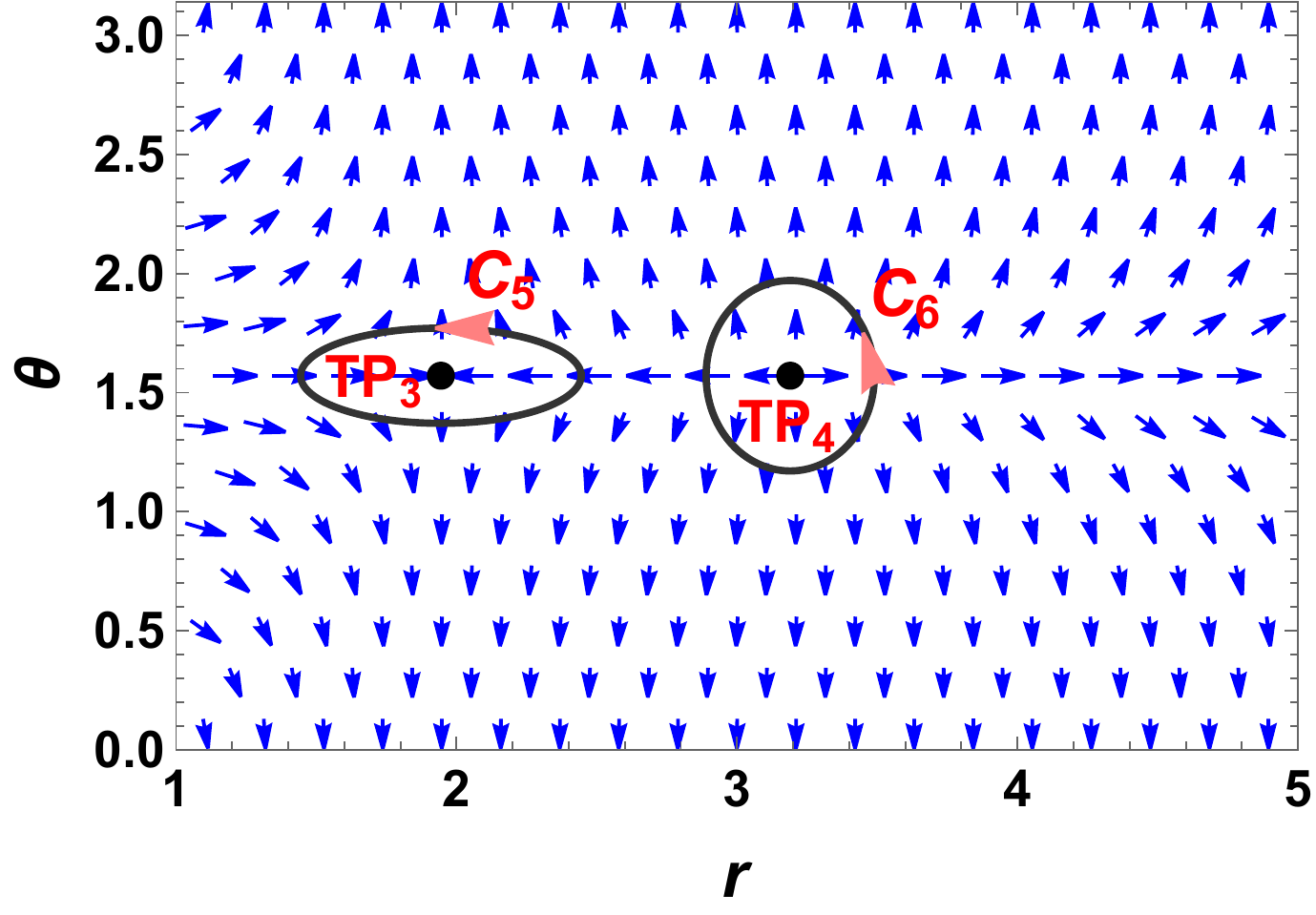}\label{emsC3N}}
    \subfigure[]{\includegraphics[width=6cm]{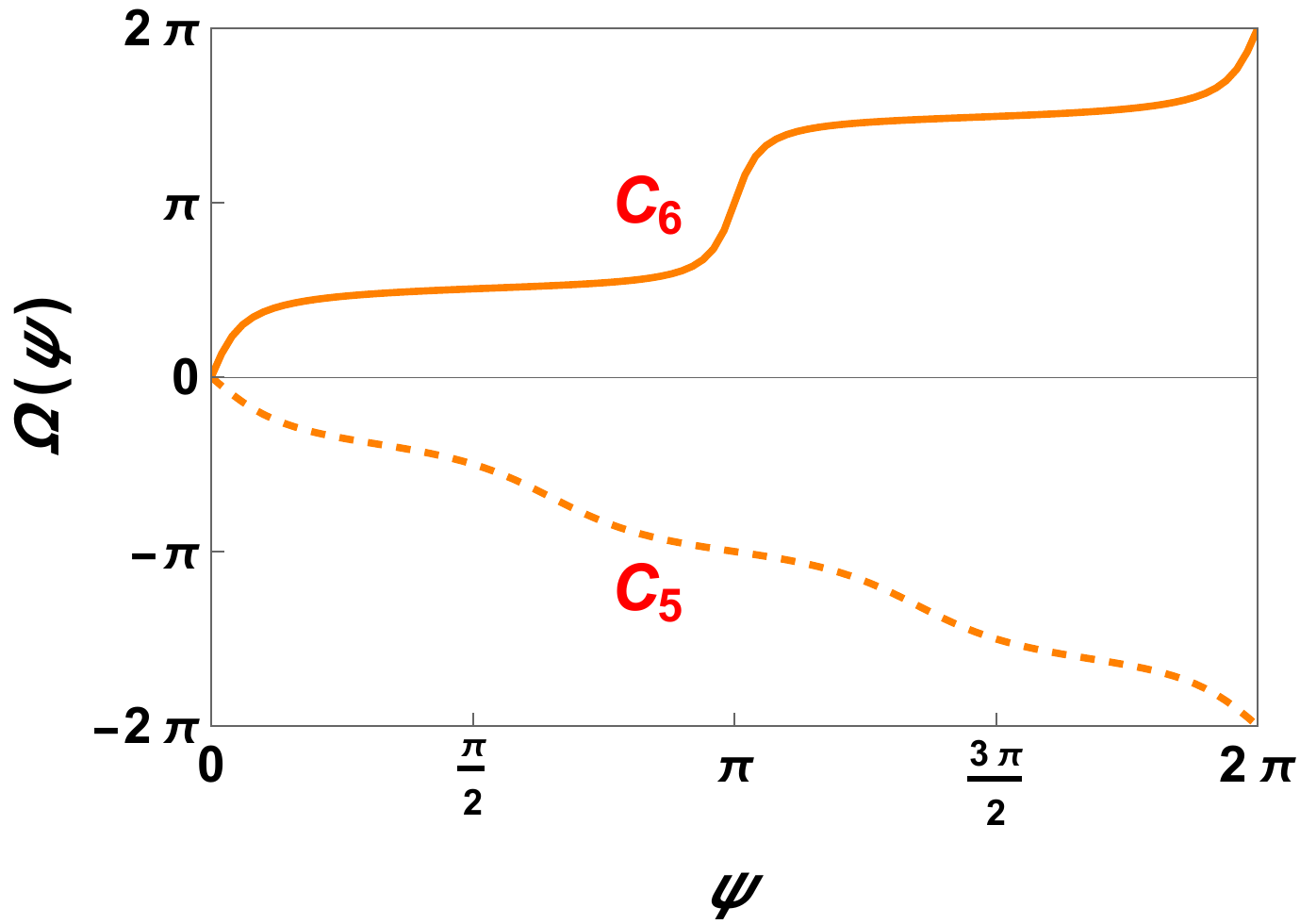}\label{emsC3A}}
    \caption{The third case with $l_{ISCO}<l=10<l_{MSCO}$ for the scalarized Einstein-Maxwell black holes. (a) The unit vector field $n$ on a portion of the $\theta$-$r$ plane. ``$TP_3$" and ``$TP_4$" denote the TCOs at $r$=1.944 and 3.189. The closed loops $C_5$ and $C_6$ have parametric coefficients ($c_0$, $c_1$, $c_2$)=(1.944, 0.5, 0.2) and (3.189, 0.3, 0.4). (b) Deflection angle \(\Omega(\psi)\) along $C_5$ and $C_6$.}
\end{figure}

For the first case $0\leq l<l_{ISCO}$, it is easy to find that there is no TCO, quite similar to that of the Schwarzschild black hole with small angular momentum, see Fig. \ref{schCase1}. Therefore the topological number $W=0$. For the second case, we set the angular momentum $l=l_{ISCO}$. Then the unit vector field $n$ and deflection angle $\Omega(\psi)$ are shown in Figs. \ref{emsC2N} and \ref{emsC2A}. From Fig. \ref{emsC2N}, one can find that the direction of the vector does not change when it crosses the ISCO point. Further by constructing the closed loop $C_4$, we observe that the winding number $w=\Omega(2\pi)$=0. When increasing the angular momentum such that $l_{ISCO}<l<l_{MSCO}$, we shall see that two TCOs emerge from the ISCO. As an example, we take $l=10$. The unit vector $n$ is displayed in Fig. \ref{emsC3N}. Two zero points $TP_3$ and $TP_4$ corresponding to the TCOs are easily observed at $r$= 1.944 and 3.189. By, respectively, constructing two closed loops $C_5$ and $C_6$, we get that the small radius TCO has $w$=-1 and the large radius TCO has $w$=+1, see Fig. \ref{emsC3A}. As expected, such pattern still gives the vanishing topological number $W=-1+1=0$, the same with that of the Schwarzschild black hole.

\begin{figure}
    \centering
    \subfigure[]{\includegraphics[width=6cm]{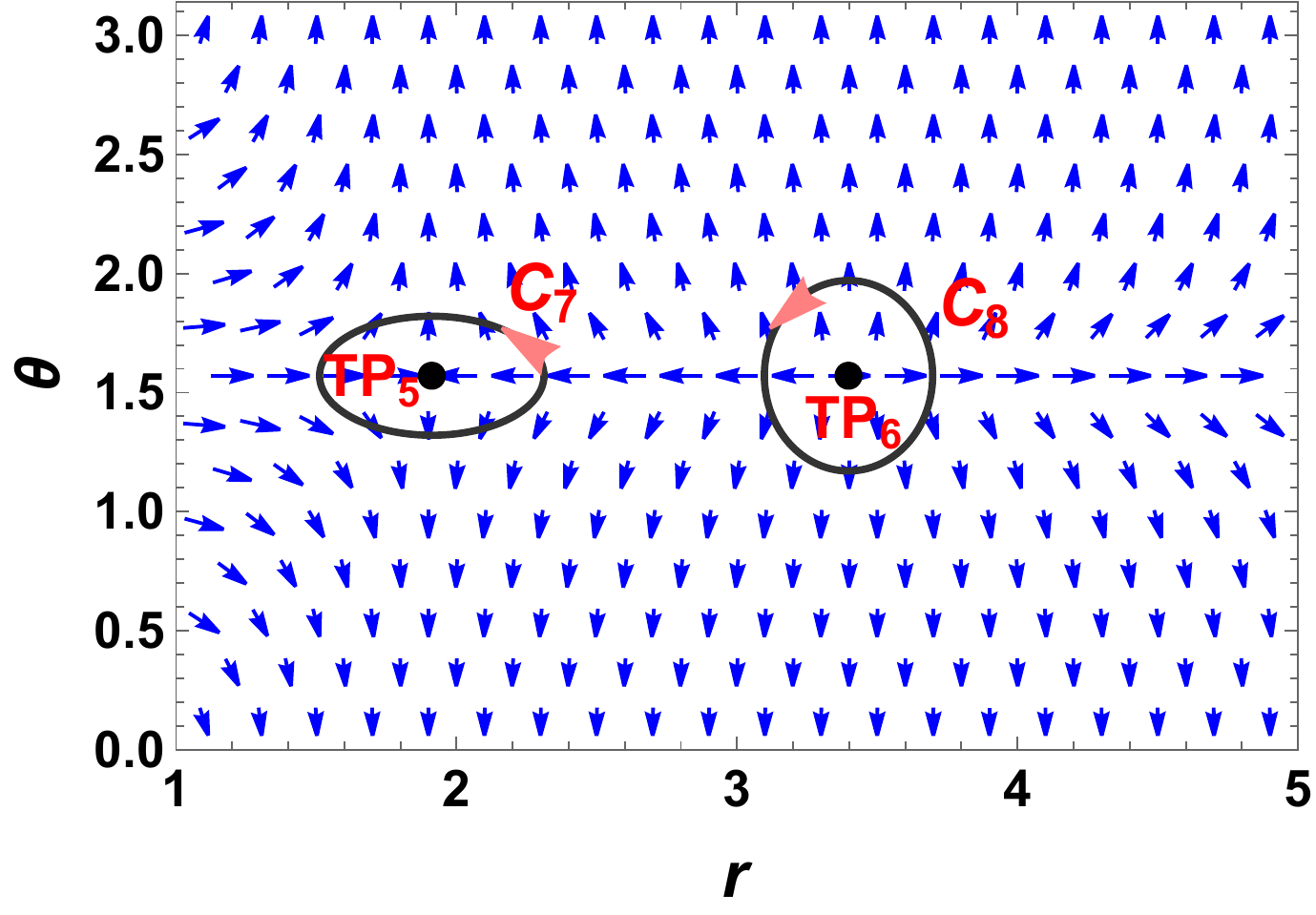}\label{emsC4N12}}
    \subfigure[]{\includegraphics[width=6cm]{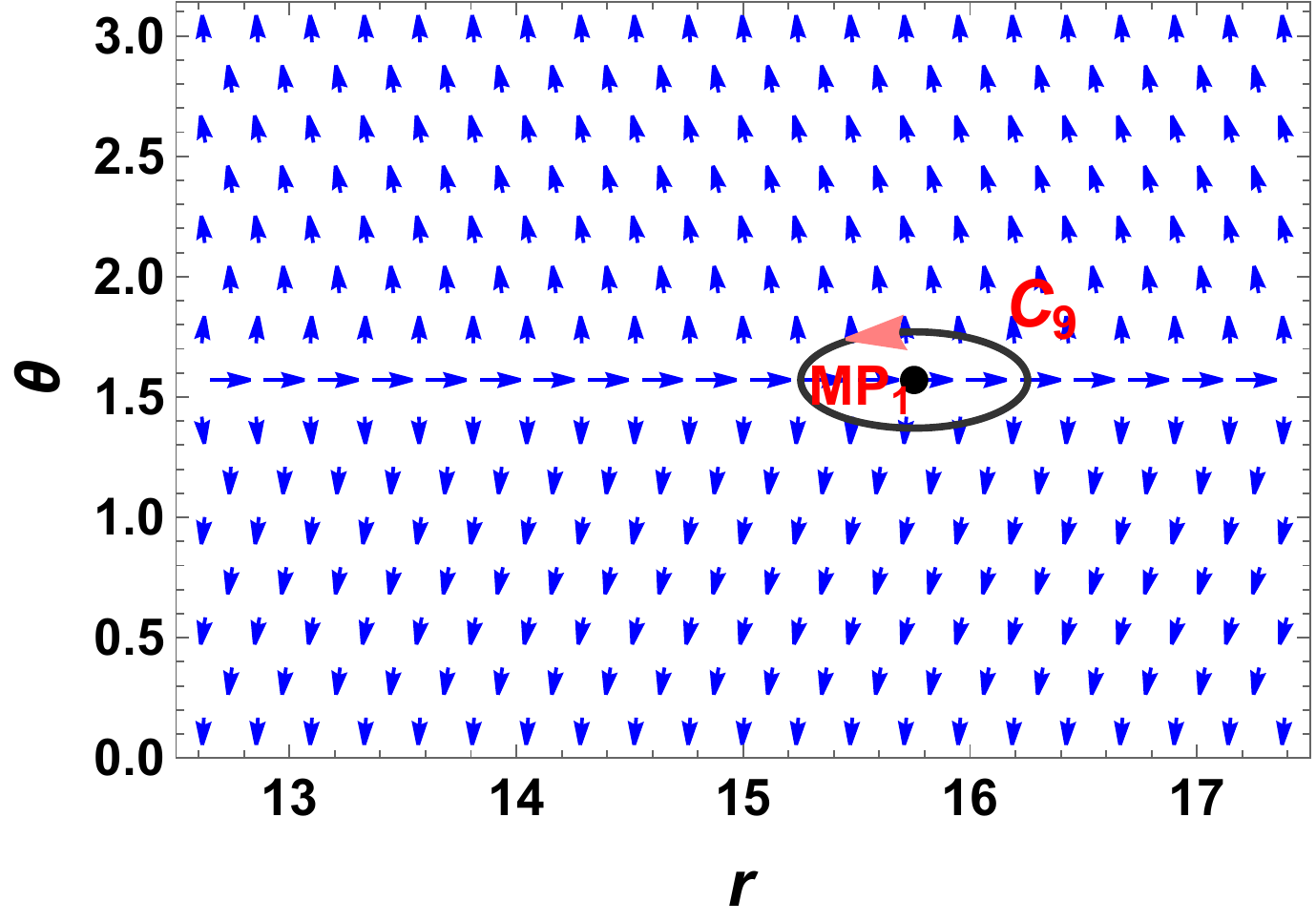}\label{emsC4Nmsco}}\\
    \subfigure[]{\includegraphics[width=6cm]{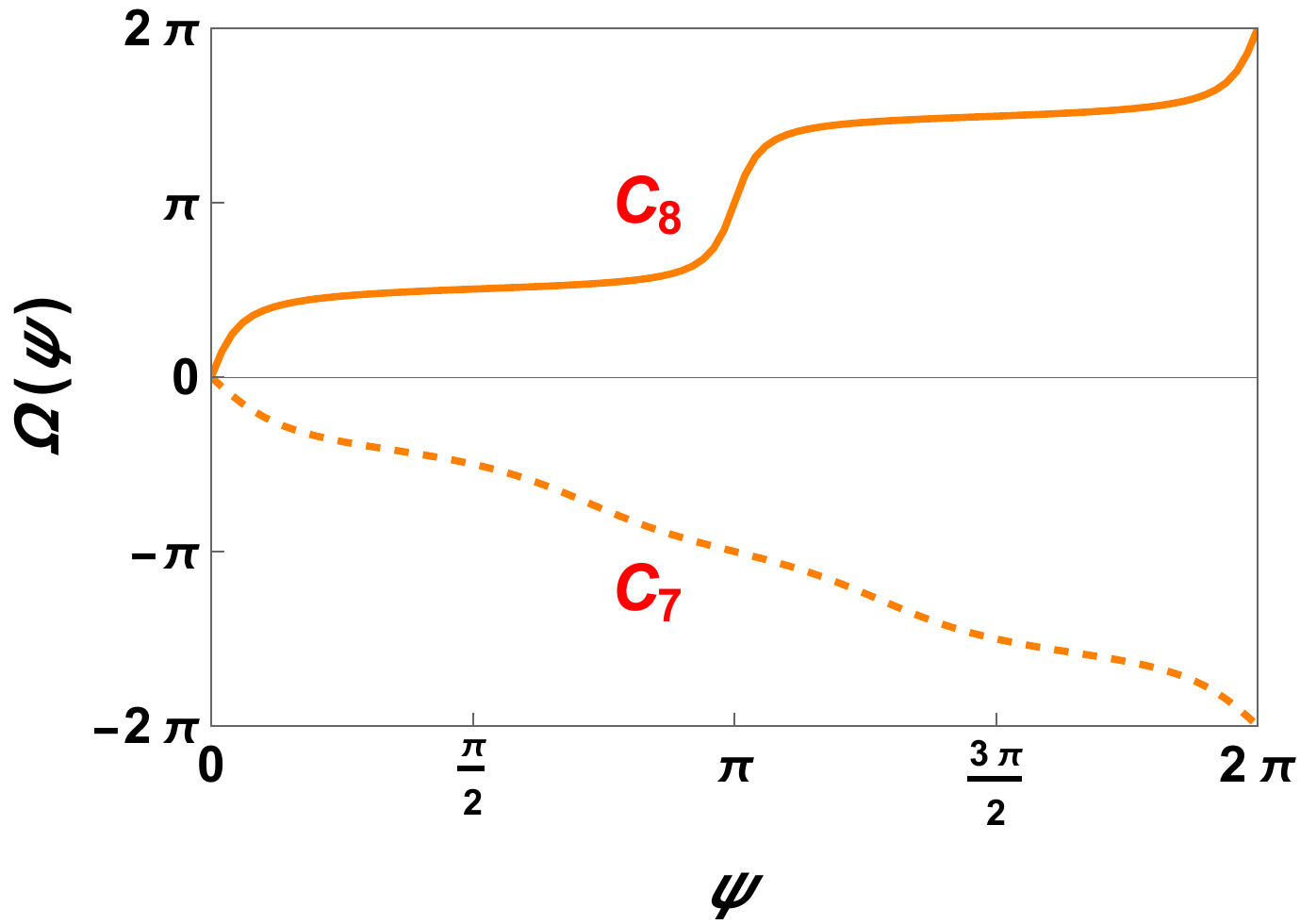}\label{emsC4A12}}
    \subfigure[]{\includegraphics[width=6cm]{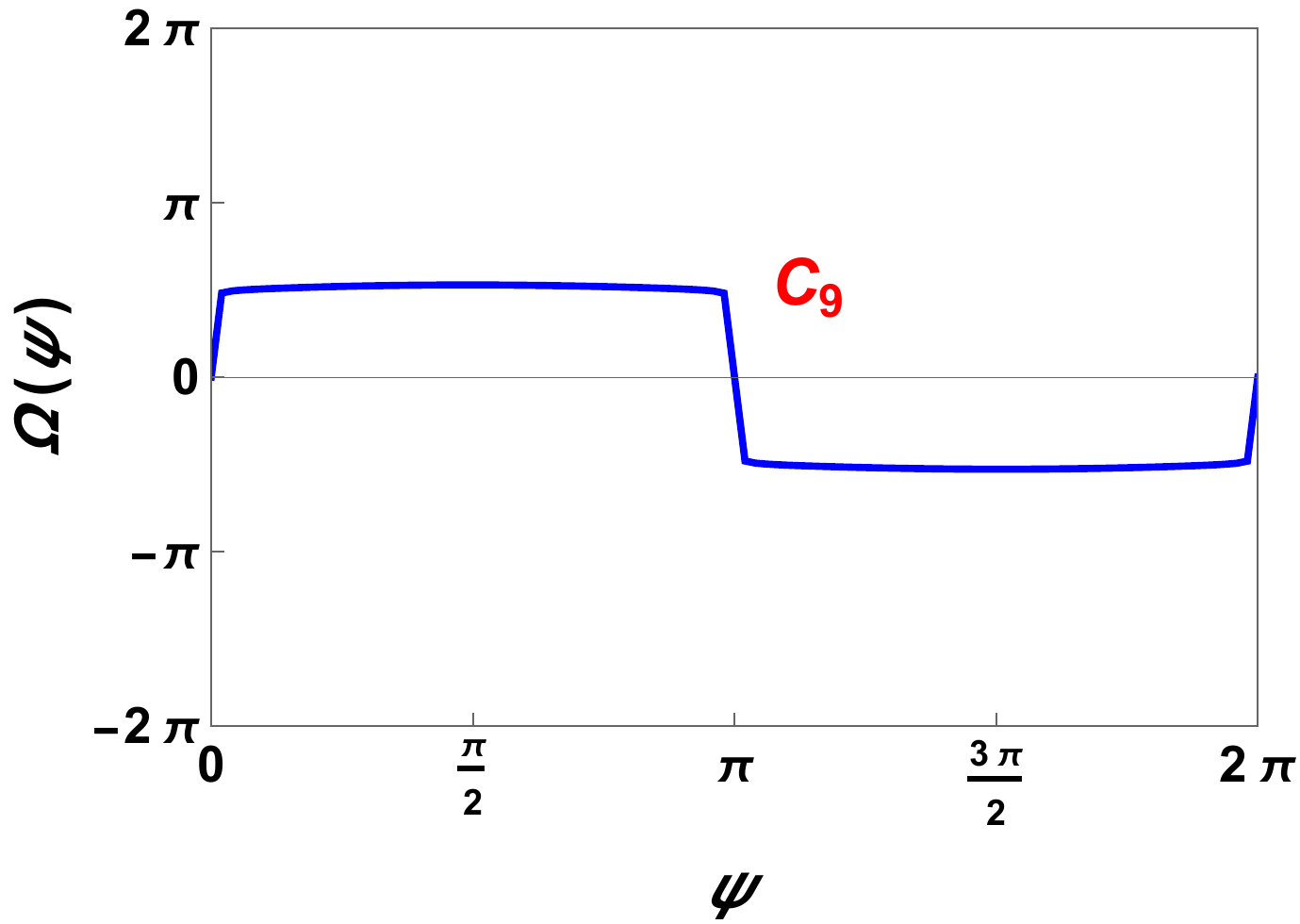}\label{emsC4Amsco}}
    \caption{The fourth case $l=l_{MSCO}$ for the scalarized Einstein-Maxwell black holes. (a) and (b) are for the unit vector field $n$ on a portion of the $\theta$-$r$ plane near the TCOs marked with $TP_5$ and $TP_6$, and MSCO with $MP_1$. The closed loops $C_7$, $C_8$, and $C_9$ are respectively, surround them. Their parametric coefficients ($c_0$, $c_1$, $c_2$)=(1.911, 0.4, 0.25), (3.394, 0.3, 0.4), and (15.753, 0.5, 0.2). (c) and (d) are for the deflection angle \(\Omega(\psi)\) along these closed loops.}
\end{figure}

Now we consider the fourth case with $l=l_{MSCO}$. The unit vector $n$ is plotted near the TCOs and MSCO in Figs. \ref{emsC4N12} and \ref{emsC4Nmsco}. Obviously, these certain orbits are the zero points of $n$ marked with black dots in the figures. Further constructing these closed loops $C_7$, $C_8$, and $C_9$, we show the deflection angle $\Omega(\psi)$ along them in Figs. \ref{emsC4A12} and \ref{emsC4Amsco}, which implies that the small and large radii TCOs denoted with $TP_5$ and $TP_6$, respectively, have $w$=-1 and 1 while MSCO's winding number vanishes. Summing up, we have the topological number $W=-1+1+0=0$, which is still the same with the previous cases.

\begin{figure}
    \centering
    \subfigure[]{\includegraphics[width=6cm]{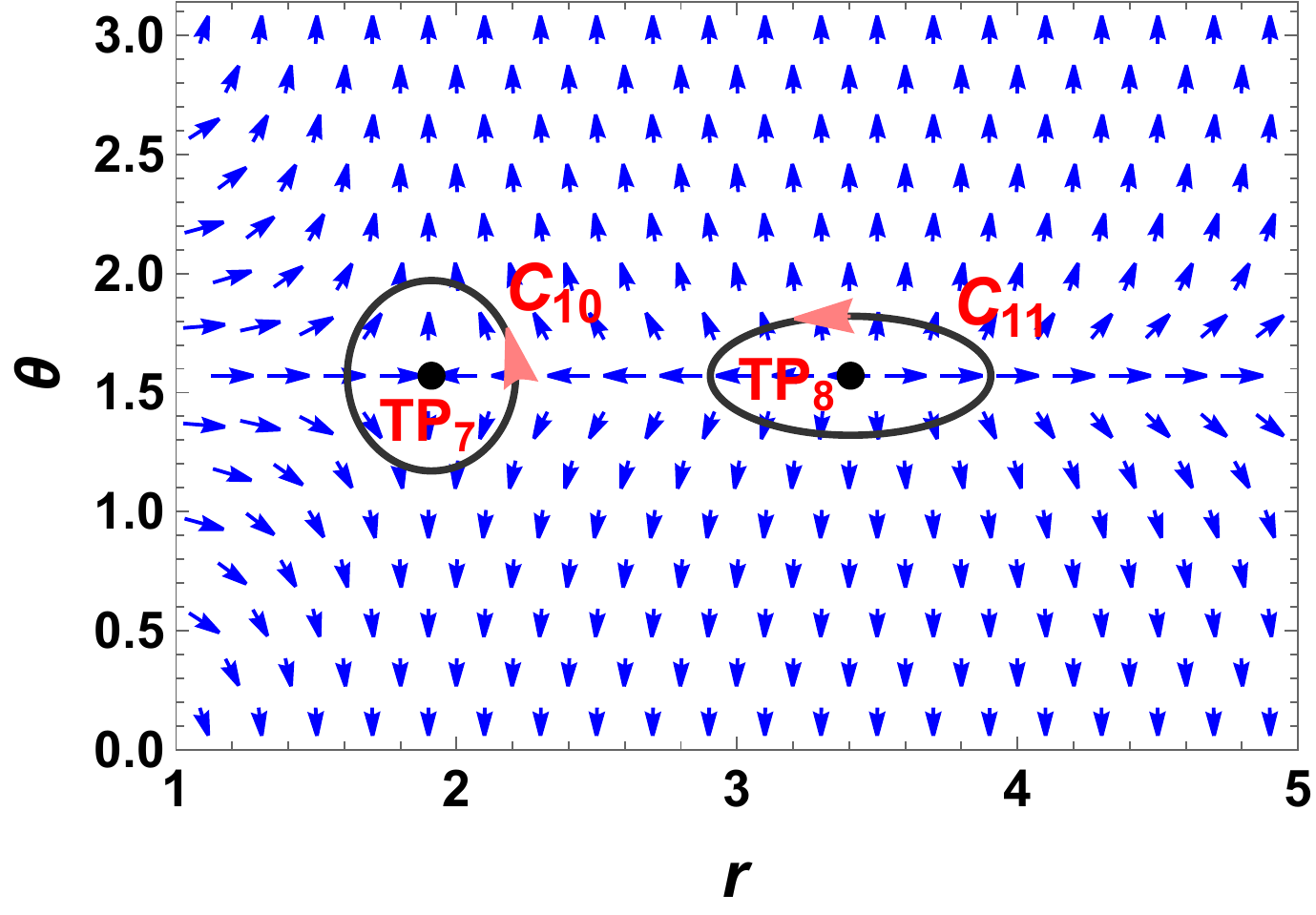}\label{emsC5N12}}
    \subfigure[]{\includegraphics[width=6cm]{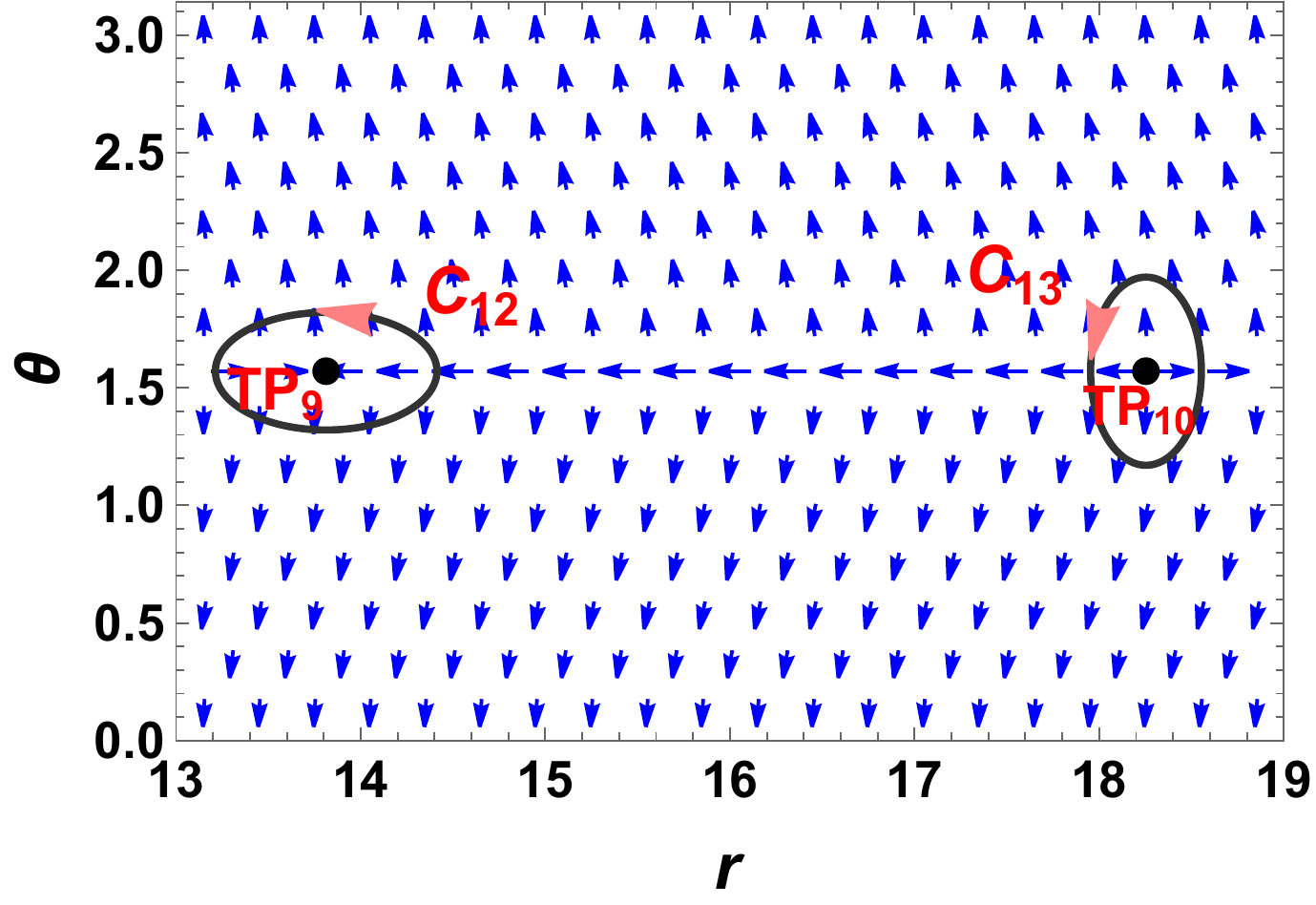}\label{emsC5N34}}
    \subfigure[]{\includegraphics[width=6cm]{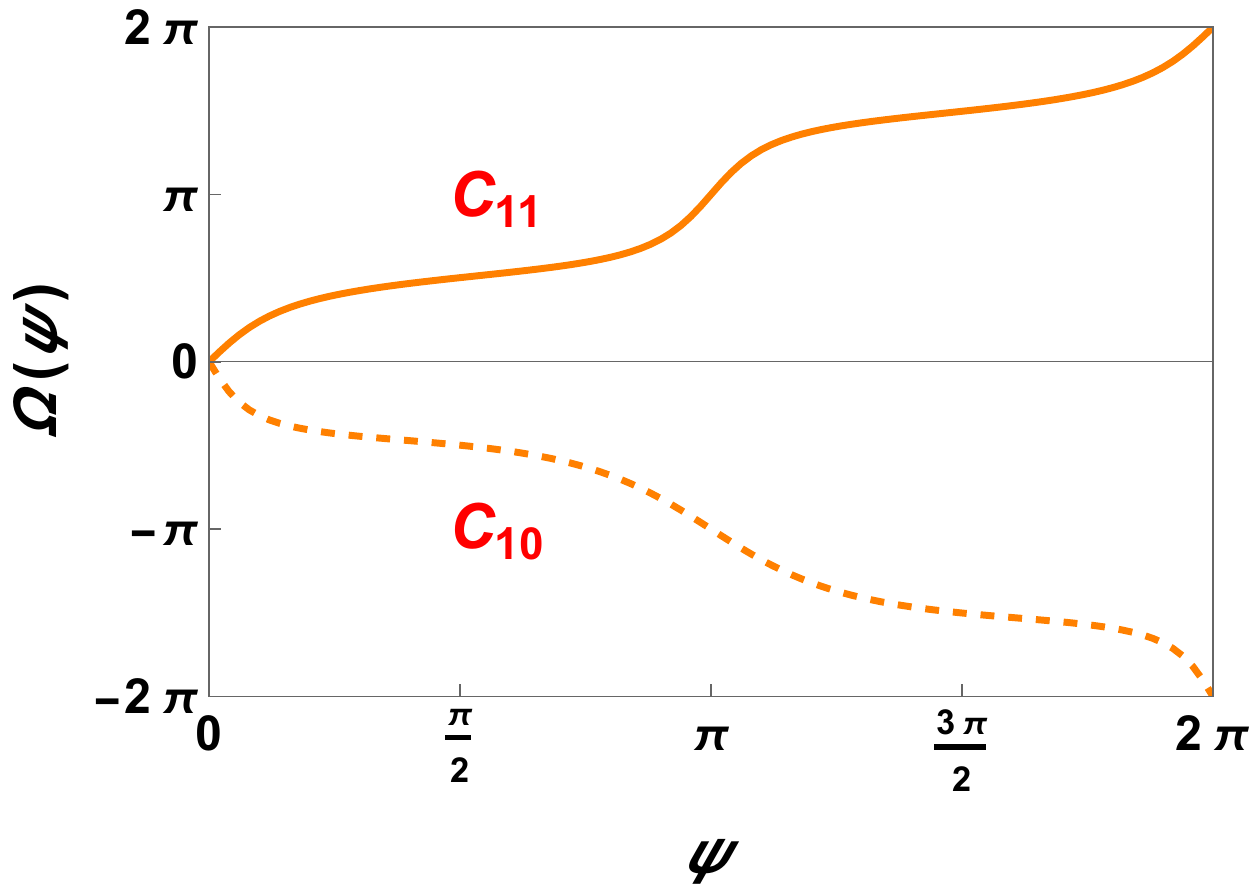}\label{emsC5A12}}
    \subfigure[]{\includegraphics[width=6cm]{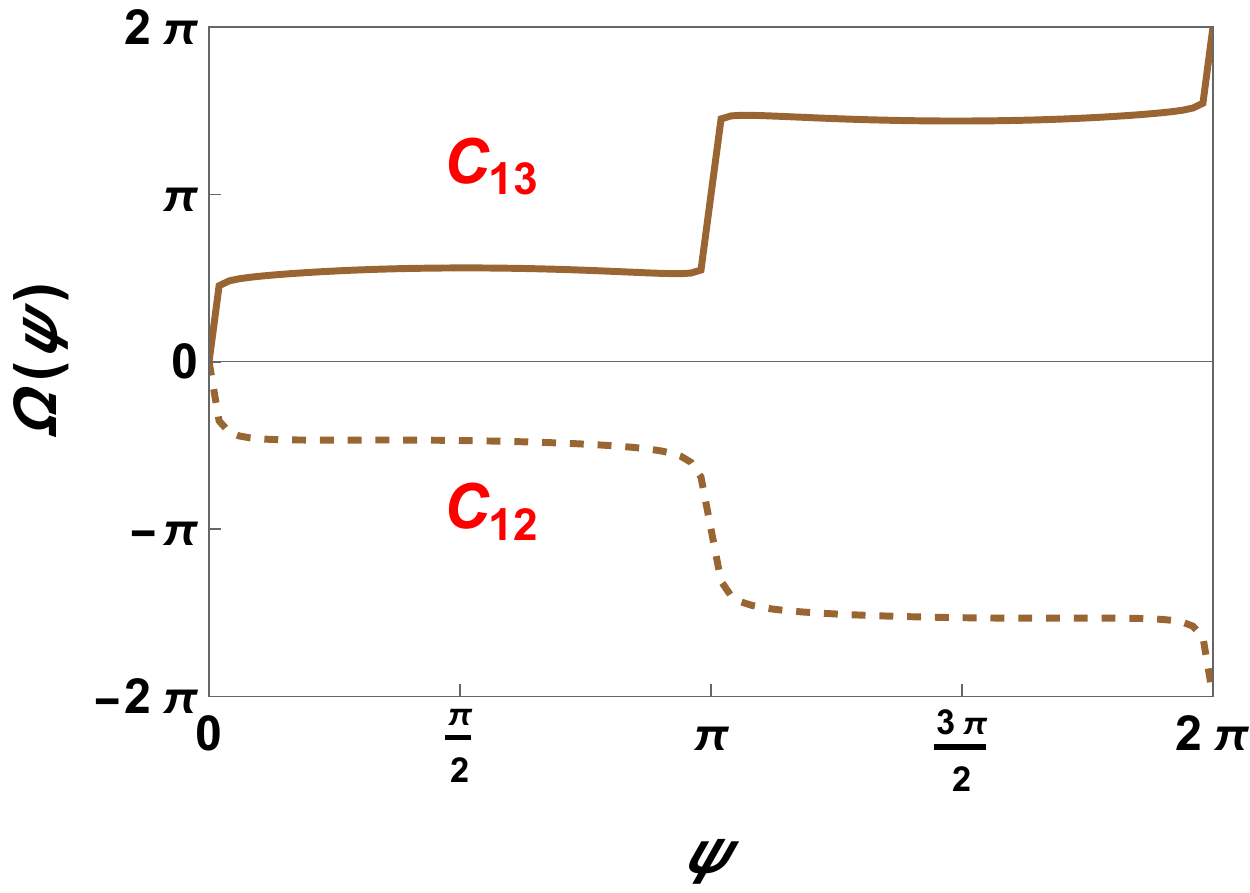}\label{emsC5A34}}
    \caption{The fifth case with $l=12.42$ for the scalarized Einstein-Maxwell black holes. (a) and (b) are for the unit vector field $n$ on a portion of the $\theta$-$r$ plane. (c) and (d) are for the deflection angle \(\Omega(\psi)\). For this case, there are four TCOs at $TP_7$, $TP_8$, $TP_9$, and $TP_{10}$. These closed loops $C_{10-13}$ have parametric coefficients ($c_0$, $c_1$, $c_2$)=(1.910, 0.3, 0.4), (3.404, 0.5, 0.25), (13.813, 0.6, 0.25), and (18.283, 0.3, 0.4).}
    \label{EMScase5}
\end{figure}

Finally, we take $l=12.42>l_{MSCO}$ for the fifth case. Similarly, we exhibit the unit vector $n$ and the deflection angle $\Omega(\psi)$ in Fig. \ref{EMScase5}. More clearly, there are four zero points located at $r$=1.911, 3.405, 13.813, and 18.253. Taking advantage of $\Omega(\psi)$, we see the winding number $w$=-1, +1, -1, and +1 for these zero points with values of $r$ from small to large. Summing these winding numbers, we have the topological number $W=-1+1-1+1=0$. This result is the same as that of the Schwarzschild black hole and our above analysis in Sec. \ref{sec2}.

\subsection{Topological configuration}

\begin{figure}
    \centering
    \subfigure[]{\includegraphics[width=6cm]{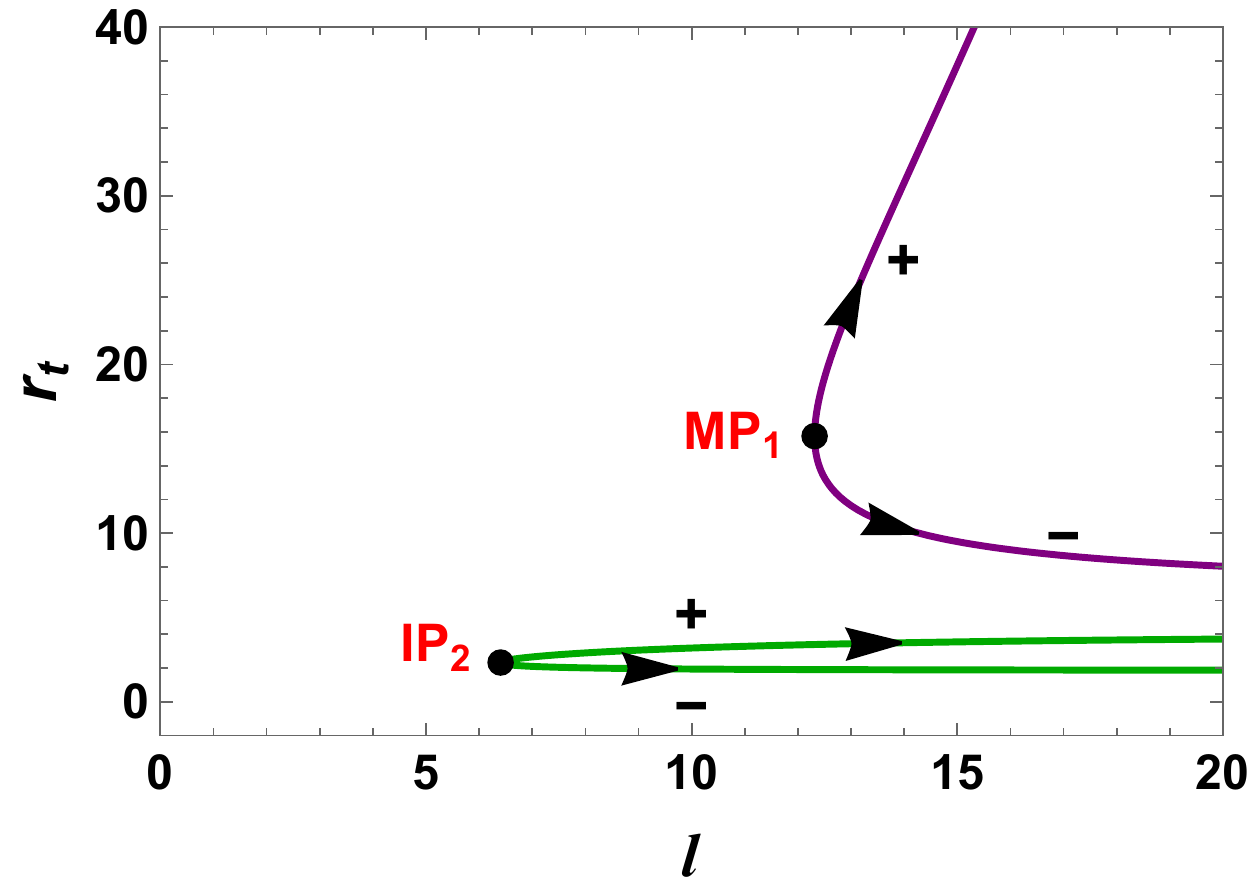}\label{EMSrL}}
    \subfigure[]{\includegraphics[width=6cm]{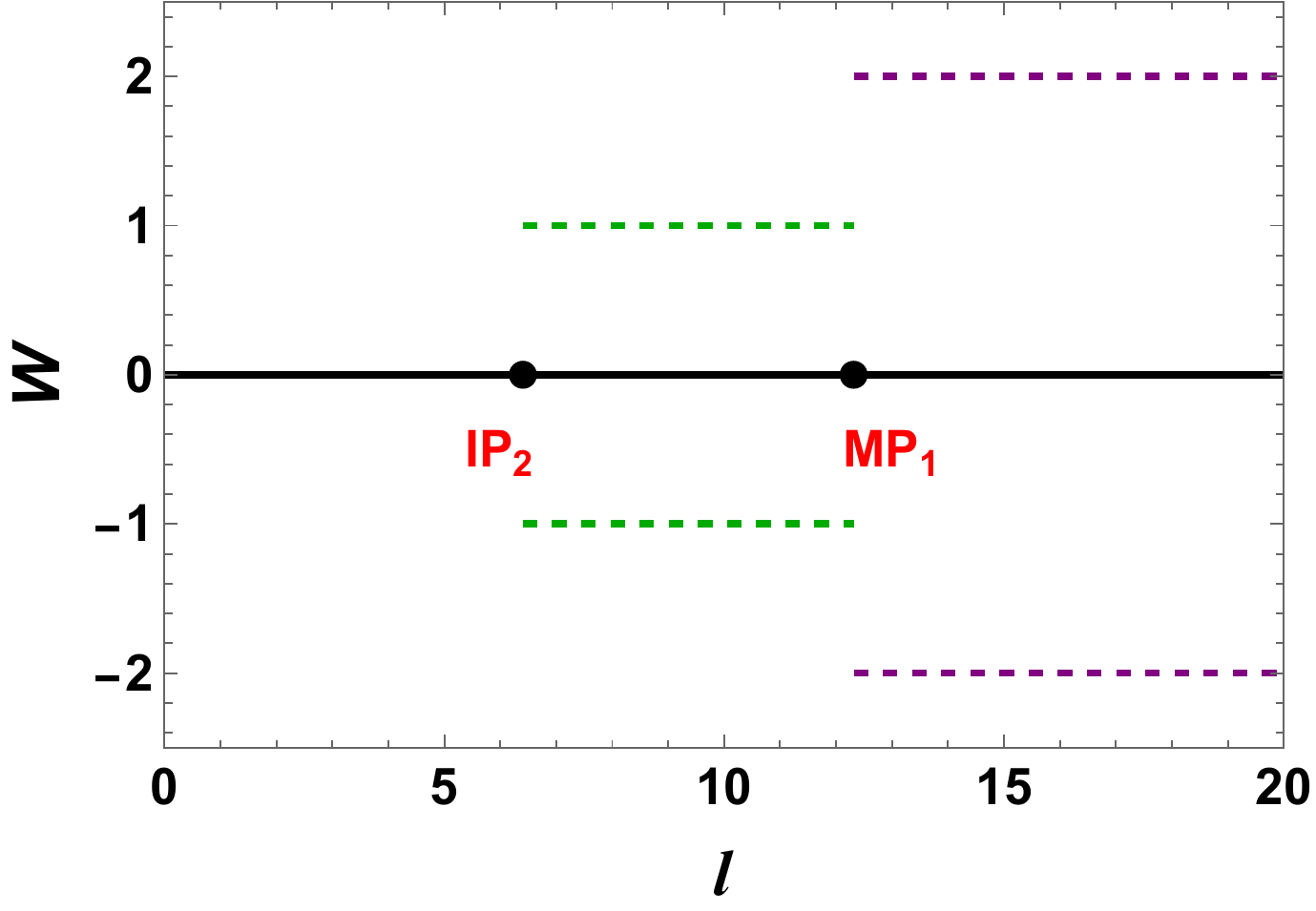}\label{EMSwL}}
    \caption{(a) The evolution of TCO radius $r_t$ vs. the angular momentum $l$ for the scalarized Einstein-Maxwell black holes. ${\rm IP}_2$ and ${\rm MP}_1$ are two generated points. The ``$\pm$" denotes that the winding number is $\pm1$ for these TCO branches. (b) The total topological number (solid black line) and winding number of TCO branches (dashed green and purple lines).}
\end{figure}

Here by considering the angular momentum as a control parameter, we depict the evolution of the zero point of the vector corresponding to the TCOs in Fig. \ref{EMSrL}. The ISCO and MSCO are marked with the black dots. Near them, we expand the angular momenta as follows
\begin{equation}
    \begin{split}
    l_t&=l_{ISCO}+7.26668 (r-r_{ISCO})^2+\mathcal{O}\left((r-r_{ISCO})^3\right), \\
    l_t&=l_{MSCO}+0.021906 (r-r_{MSCO})^2+\mathcal{O}\left((r-r_{MSCO})^3\right).
    \end{split}
\end{equation}
Obviously, both $l_t''(r_{ISCO})$ and $l_t''(r_{MSCO})$ are positive, which suggests that both the bifurcation points are generated points. It is easy to see that, near each of them, two TCO branches emerge, one of which has positive winding number $w=1$ while another has negative winding number $w=-1$ marked with ``+" and ``-" in the figure. Nevertheless, the topological number $W$ always vanishes.

Meanwhile, the winding number with respect to the angular momentum is illustrated in Fig. \ref{EMSwL}. For \(l_{ISCO}<l<l_{MSCO}\), two TCOs with opposite winding numbers are described as a green line, resulting in $W=0$. When $l>l_{MSCO}$, Four TCOs appear, two of them are stable and have $W=+2$, whereas the other two unstable ones have $W=-2$, thus their total winding number is still zero.

In summary, we find that there will be no, one pair, or two pairs TCOs for different values of the angular momentum, which shows a new topological configuration different from the Schwarzschild and Kerr black holes. Despite this fact, the topological number still keeps zero. So different configurations of TCOs could give the same topological number and belong to the same topological class.

\section{Dyonic Black Holes}\label{sec5}

As shown above, for the scalarized Einstein-Maxwell black holes, the MSCO and ISCO are present, which gives a different topology configuration of TCOs from that of the Schwarzschild black holes, see Figs. \ref{EMSrL} and \ref{schRL}. In this section, we would like to exhibit another characteristic topological configuration of TCOs, where the ISCO needs not to satisfy the condition $\partial_{r,r}e_1(r)=0$. For this characteristic case, we also expect to check whether the topological number $W$ still vanishes.

\subsection{Dyonic black holes}

Here we focus on the dyonic black holes with the quasi-topological electromagnetism, while it has no contributions to the Maxwell equation and energy-momentum tensor.

The Lagrangian with the quasi-topological electromagnetism is written as \cite{Liu2019}
\begin{equation}\label{dyLag}
    \mathcal{L}=\sqrt{-g}(R+\alpha_1 U^{(1)}-\alpha_2 U^{(2)}),
\end{equation}
where \(\alpha_1\) and \(\alpha_2\) are coupling constants. \(U^{(1)}=-F^2\) is the conventional Maxwell Lagrangian and \(U^{(2)}=-2F^4+(F^2)^2\) is a quasi-topological electromagnetism term. Here $F^2=F^{\mu\nu}F_{\mu\nu}$ and $F^4=F^\mu_{\;\nu} F^\nu_{\;\rho} F^\rho_{\;\sigma} F^\sigma_{\;\mu}$. From the Lagrangian (\ref{dyLag}), the Bianchi identity and Maxwell equation of motion read
\begin{equation}
    \begin{split}
    &\text{BI: } \nabla_{[\mu}F_{\nu\rho]}=0,\quad \text{EOM: }\nabla_\mu\tilde{F}^{\mu\nu}=0,\\
    & \tilde{F}^{\mu\nu}=4\alpha_1 F^{\mu\nu}+8\alpha_2(F^2 F^{\mu\nu }-2F^{\mu\rho }F^\sigma_{\;\rho} F_\sigma^{\;\nu}).
    \end{split}
\end{equation}
Simultaneously, the Einstein field equations are
\begin{equation}
    \begin{split}
    &R_{\mu\nu}-\frac{1}{2}g_{\mu\nu}=T_{\mu\nu},\\
    &T_{\mu\nu}=\alpha_1 (2F_{\mu\rho}F_\nu^{\;\rho}-\frac{1}{2}F^2g_{\mu\nu})+\alpha_2(4F^2F_{\mu\rho}F_\nu^{\;\rho}-8F_{\mu\rho} F^\rho_{\;\sigma} F^\sigma_{\;\lambda} F^\lambda_{\;\nu} -\frac{1}{2}((F^2)^2-2F^4)g_{\mu\nu}).
    \end{split}
\end{equation}
The quasi-topological electromagnetism admits spherically symmetric dyonic black hole solution described by the following line element \cite{Liu2019}
\begin{align}
    ds^2 & =-f(r) dt^2+ \frac{1}{f(r)}dr^2+r^2 d\Omega^2, \\
    f(r) & =1-\frac{2M}{r}+\frac{\alpha_1 p^2}{r^2 }+\frac{q^2}{\alpha_1 r^2} {\;}_2F_1\left( \frac{1}{4},1;\frac{5}{4};-\frac{4p^2\,\alpha_2}{r^4\, \alpha_1}\right), \label{dymetric}
\end{align}
where $p$ and $q$ correspond to the magnetic charge and electric charge of the black holes. \({\;}_2F_1\) is the hypergeometric function. Coupling constants \(\alpha_1\) and \(\alpha_2\) are associated with Maxwell theory and quasi-topological electromagnetism term, respectively. For a characteristic case, we set \(\alpha_1=1\), $q=6.85$, $p=\sqrt{\frac{396}{443}}$, $\alpha_2=\frac{196249}{1584}$, and $M=6.7$. Accordingly, the effective potential reads
\begin{equation}
    \mathcal{V}(r)=\frac{-\alpha_1 E^2 r^2}{ _2F_1\left(\frac{1}{4},1;\frac{5}{4};-\frac{4 p^2 \alpha_2}{r^4 \alpha_1}\right) q^2+\alpha_1 \left(\alpha_1 p^2+r^2-2 M r\right)}+\frac{l^2 \csc ^2\theta}{r^2}+\mu ^2,
\end{equation}
by making use of Eq. (\ref{Vr}).

\subsection{Topology of TCOs and winding number}

In order to investigate the topology of the TCO for the dyonic black hole, we evaluate $e_1$ and $e_2$ via Eq. (\ref{e1e2})
\begin{equation}
    e_{1,2}=\pm\sqrt{\frac{\left(l^2 \csc ^2\theta+\mu ^2 r^2\right) \left( _2F_1\left(\frac{1}{4},1;\frac{5}{4};-\frac{4 p^2 \alpha _2}{r^4 \alpha_1}\right)q^2 +\alpha _1 \left(-2 M r+\alpha _1 p^2+r^2\right)\right)}{\alpha_1\; r^4}}.
\end{equation}
Based on them, the components \(\phi^r\) and \(\phi^\theta\) of the vector can be easily calculated
\begin{align}
    \phi^r= & \frac{q^2 \csc^2\theta(-6l^2+r^2\csc(2\theta)-r^2)(4 \alpha_2 p^2+\alpha_1 r^4)\; _2F_1\left(\tfrac{1}{4},1;\tfrac{5}{4};\tfrac{-4p^2 \alpha_2}{r^4 \alpha_1}\right)}{4 \alpha_1 r^4(4 \alpha_2 p^2+\alpha_1 r^4) \sqrt{l^2 \csc^2\theta+ r^2}}  \nonumber\\
    & \quad -\frac{2 \alpha_1l^2 \csc^2\theta (q^2 r^4-2(3M r-2 \alpha_1 p^2 -r^2)(4 \alpha_2 p^2+\alpha_1 r^4))}{4 \alpha_1 r^4(4 \alpha_2 p^2+\alpha_1 r^4) \sqrt{l^2 \csc^2\theta+ r^2}}\\
    & \quad -\frac{2 \alpha_1 r^2(q^2 r^4-2 (M r-\alpha_1 p^2)(4 \alpha_2 p^2 +\alpha_1 r^4))}{4 \alpha_1 r^4(4 \alpha_2 p^2+\alpha_1 r^4) \sqrt{l^2 \csc^2\theta+ r^2}},\nonumber \\
    \phi^\theta= & \frac{-l^2\cot \theta \csc^2 \theta \sqrt{\alpha_1 (l^2 \csc^2\theta +r^2)(_2F_1\left(\tfrac{1}{4},1;\tfrac{5}{4};\tfrac{-4p^2 \alpha_2}{r^4 \alpha_1}\right) q^2+ \alpha_1(-2Mr +\alpha_1 p^2+r^2))}}{\alpha_1 l^2 r^3 \csc^2\theta +\alpha_1 r^5}.
\end{align}
Adopting Eq. (\ref{iscophi}), the radius and angular momentum of MSCO for dyonic black hole is
\begin{equation}
    r_{MSCO}=25.3799, \quad l_{MSCO}=18.52.
\end{equation}
Note that this MSCO is not the innermost TCO, so the ISCO does not coincide with this MSCO. The actual ISCO locates at $r_{ISCO}=6.0928$ with $l_{ISCO}=0$ and $E_{ISCO}=0.1393$.

According to the angular momentum of the MSCO and ISCO, we can divide them into following three characteristic cases,
\begin{itemize}
    \item \( 0\leq l<l_{MSCO} \),
    \item \(  l=l_{MSCO} \),
    \item \( l_{MSCO}<l<\infty \).
\end{itemize}
For the first case with small angular momentum, the study shows that there are two TCOs. This is quite different from that of the Schwarzschild black holes and scalarized Einstein-Maxwell black holes, where no TCOs can be found. This phenomenon is mainly caused by the quasi-topological electromagnetism term. More interestingly, they are not generated from an ISCO. Taking $0<l=2<l_{MSCO}$ as an example, the radius of TCOs locates at
\begin{equation}
    r_{TP_{11}}=2.63492,  \quad  r_{TP_{12}}=6.11503.
\end{equation}
Their stability is evaluated via the first derivative
\begin{equation}
    \phi^{r\prime}(r_{TP_{11}})=-0.07678,  \quad  \phi^{r\prime}(r_{TP_{12}})=0. 01462,
\end{equation}
which significantly implies that $TP_{11}$ is unstable while $TP_{12}$ is stable.

For the second case $l=l_{MSCO}$, we find the vector satisfies
\begin{equation}
    \phi^{r\prime}(r_{MSCO})=0,\quad \phi^{r\prime\prime}(r_{MSCO})=0.00001835.
\end{equation}
at $r_{MSCO}$, which indicates that MSCO obeys the condition (\ref{SCOMCO}). For the third case, there will be four TCOs with two being stable and other two unstable.

\begin{figure}[h]
    \centering
    \subfigure[]{\includegraphics[width=6cm]{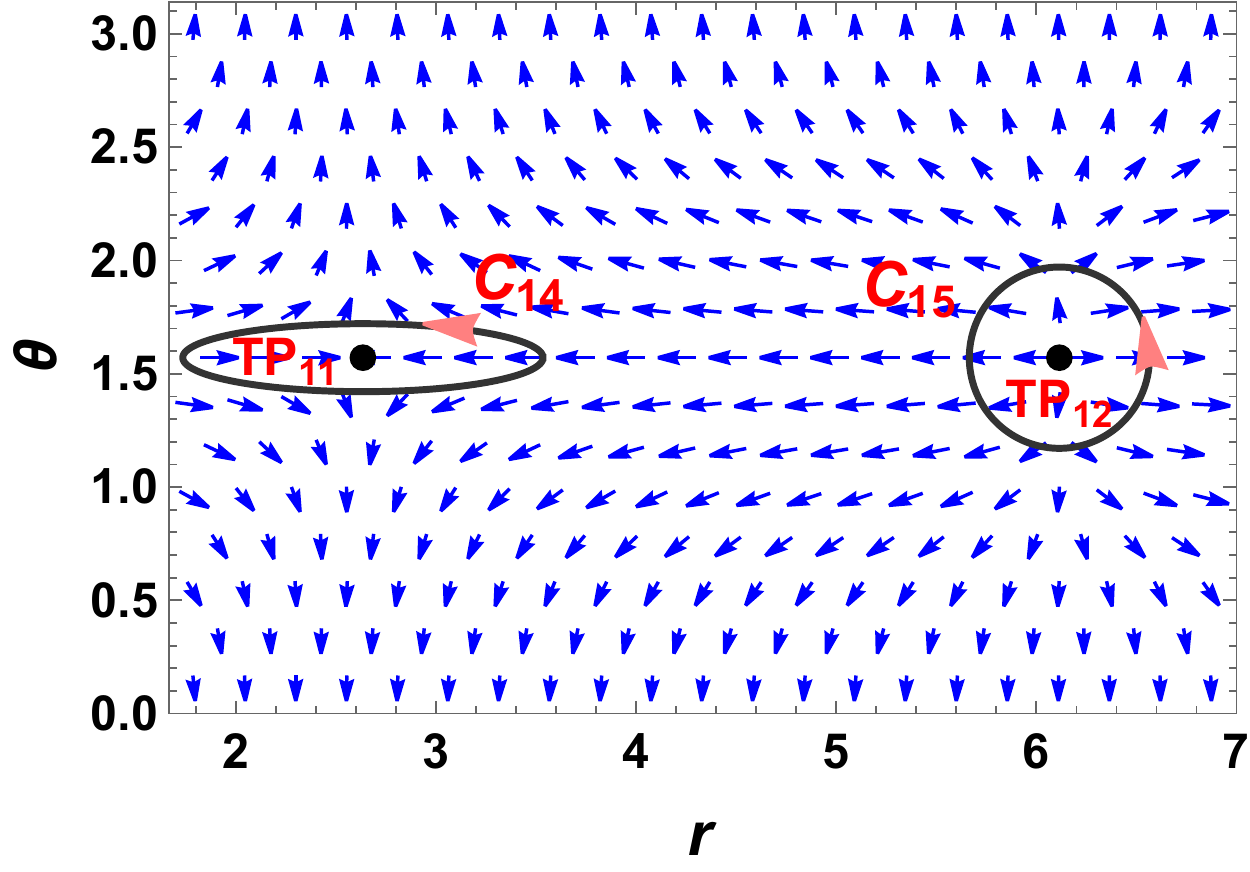}\label{dyC1N}}
    \subfigure[]{\includegraphics[width=6cm]{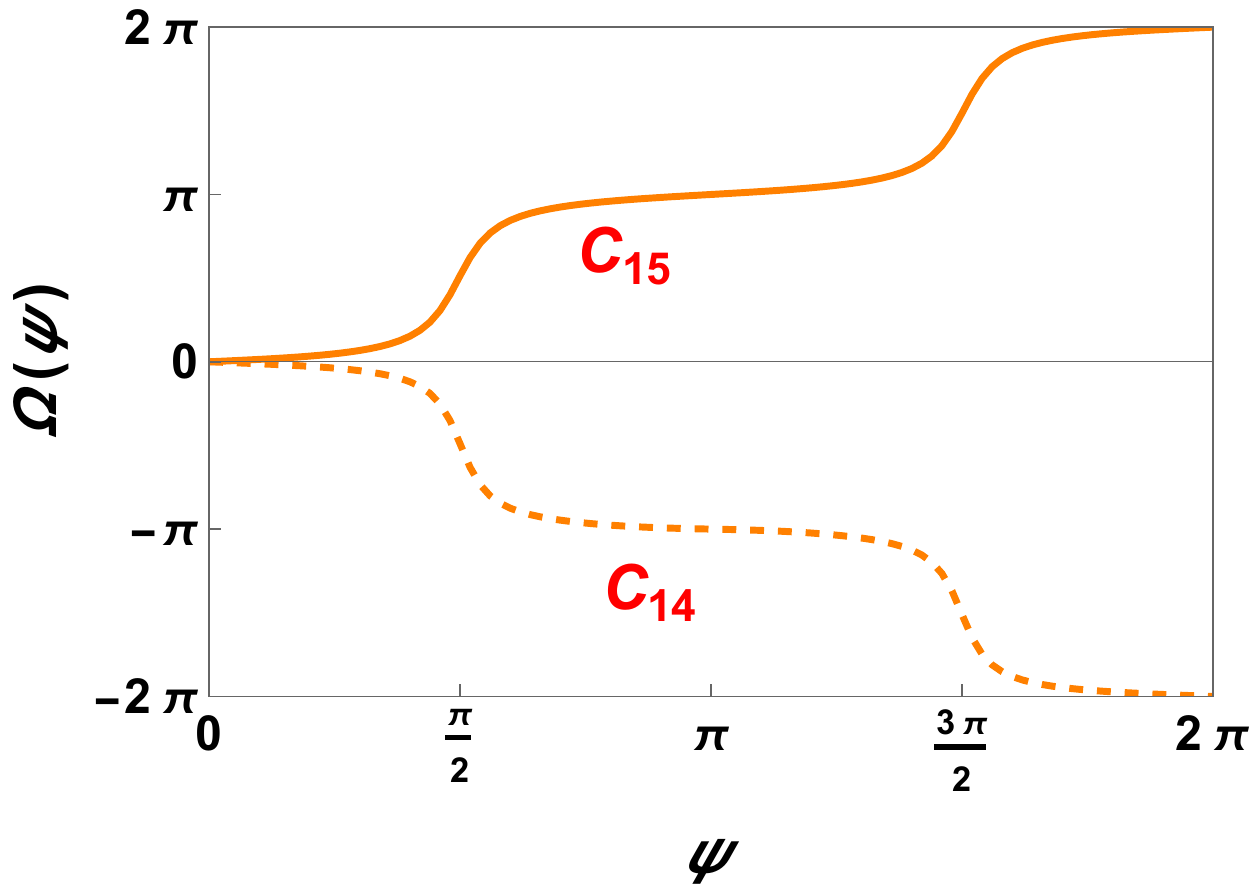}\label{dyC1A}}
    \caption{The first case with $l=2.0$ for the dyonic black holes. (a) The unit vector field $n$ on a portion of the $\theta$-$r$ plane. ``$TP_{11}$" and ``$TP_{12}$" are two TCOs at $r$=2.635 and 6.115. The closed loops $C_{14}$ and $C_{15}$ have parametric coefficients ($c_0$, $c_1$, $c_2$)=(2.635, 0.9, 0.15) and (6.115, 0.45, 0.4). (b) Deflection angle \(\Omega(\psi)\) along $C_{14}$ and $C_{15}$.}\label{dyC1Nad}
\end{figure}

Next, we turn to the topology for the TCOs. For the first case with $l=2$, we observe that there are two zero points of the unit vector field \(n=(n^r, n^\theta)\) marked with black dots in Fig. \ref{dyC1N}. By constructing two closed loops $C_{14}$ and $C_{15}$, we calculate the deflection angle $\Omega(\psi)$, and plot them in Fig. \ref{dyC1A}. Along $C_{14}$ or $C_{15}$, $\Omega(\psi)$ decreases to $-2\pi$ or increases to 2$\pi$. This result suggests that the winding number of $TP_{11}$ and $TP_{12}$ are -1 and +1, respectively. Summing them, one obtains the topological number $W=-1+1=0$. This result indicates that although the quasi-topological electromagnetism term produces two new TCOs, the topological number stays unchanged.

\begin{figure}
    \centering
    \subfigure[]{\includegraphics[width=6cm]{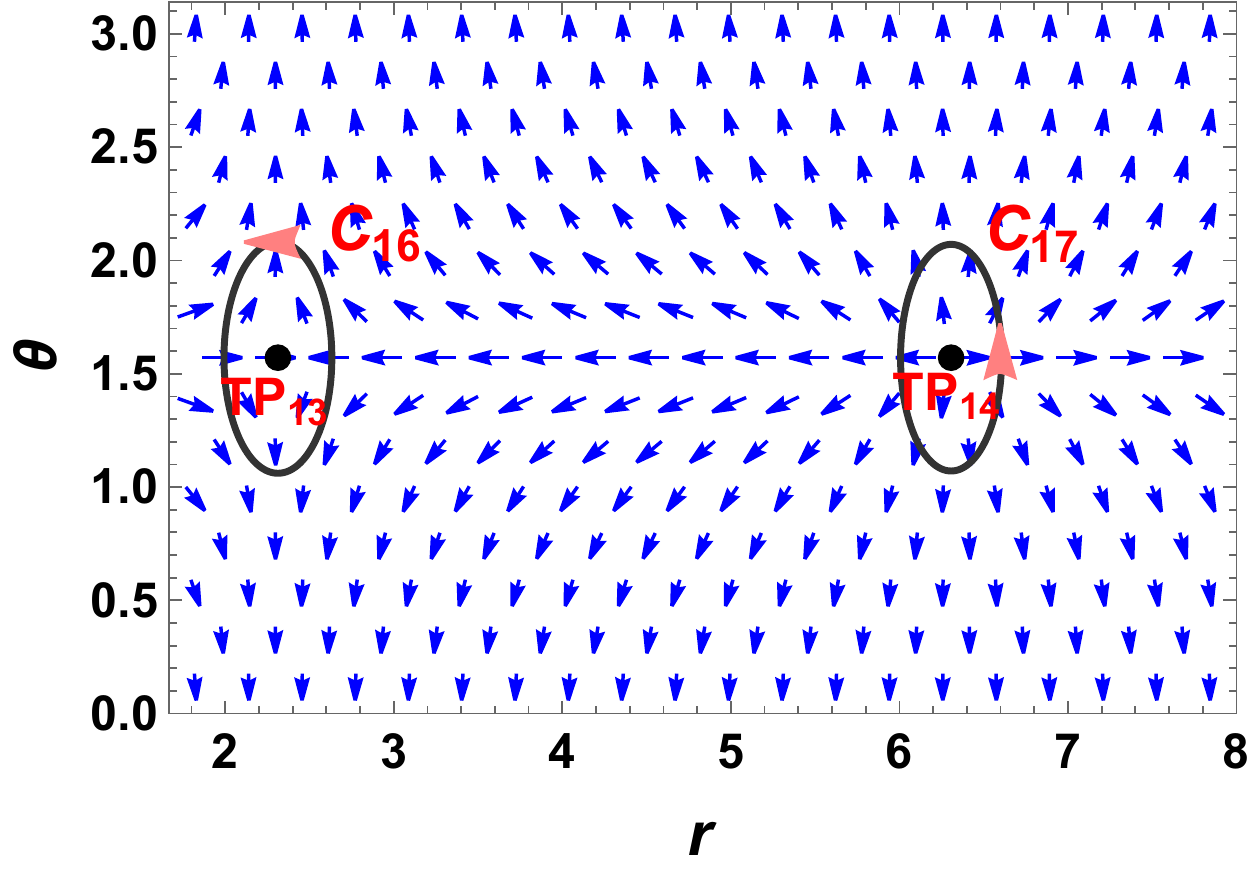}\label{dyC2N12}}
    \subfigure[]{\includegraphics[width=6cm]{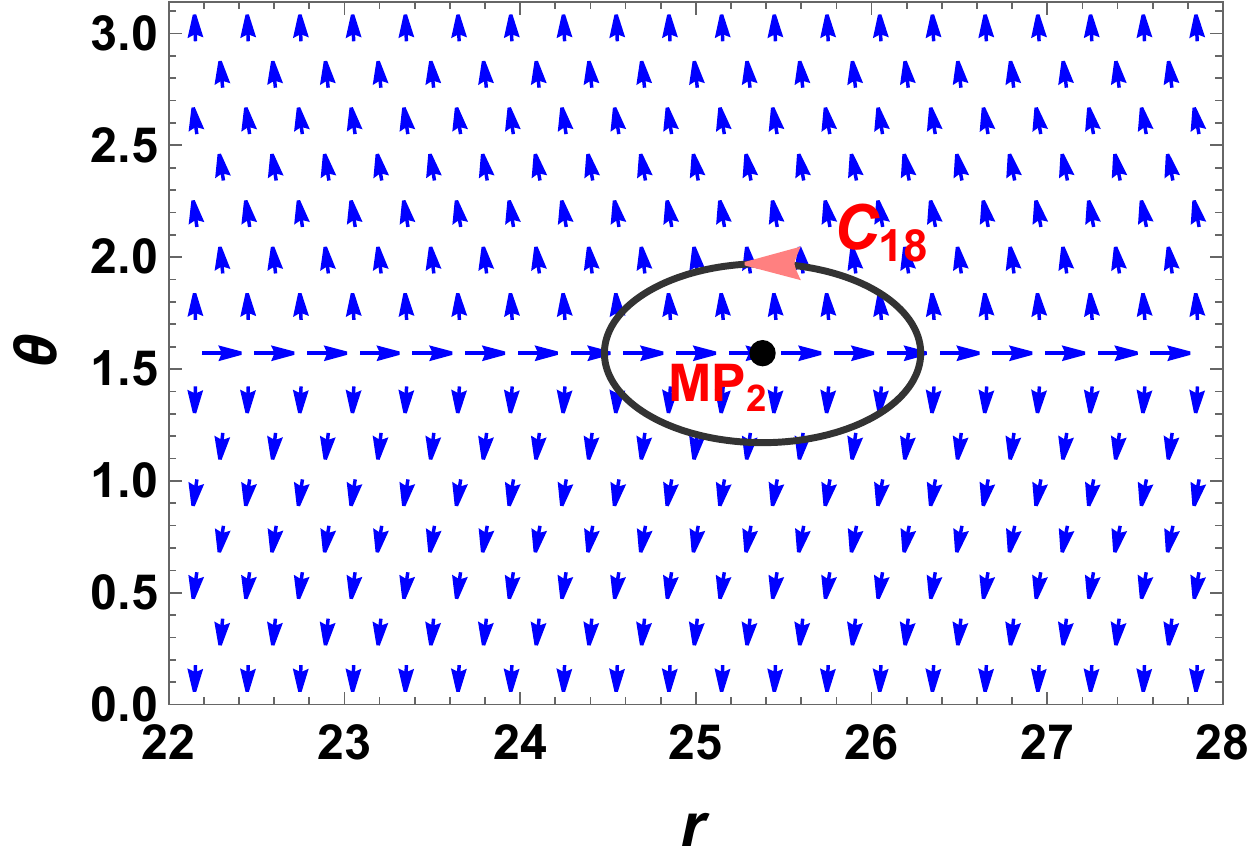}\label{dyC2Nmsco}}\\
    \subfigure[]{\includegraphics[width=6cm]{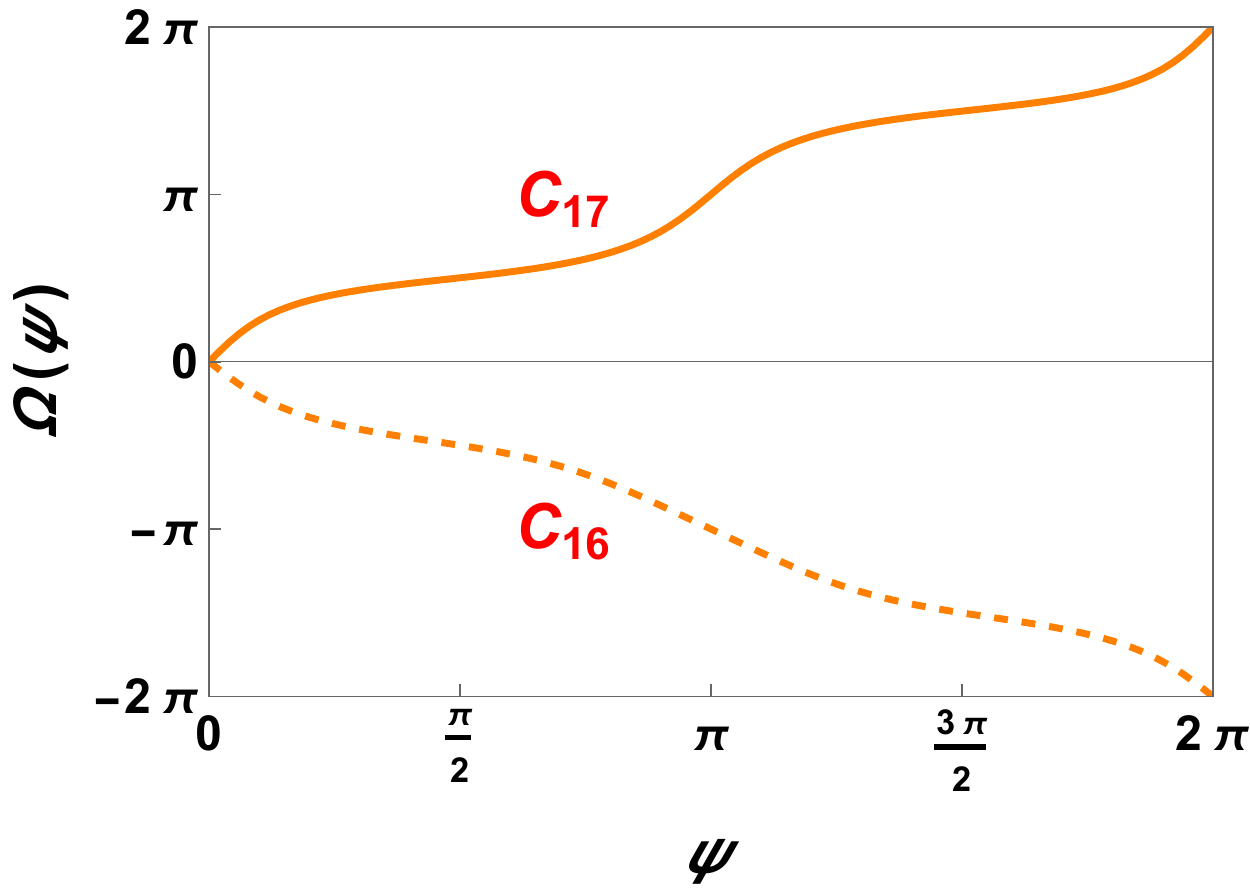}\label{dyC2A12}}
    \subfigure[]{\includegraphics[width=6cm]{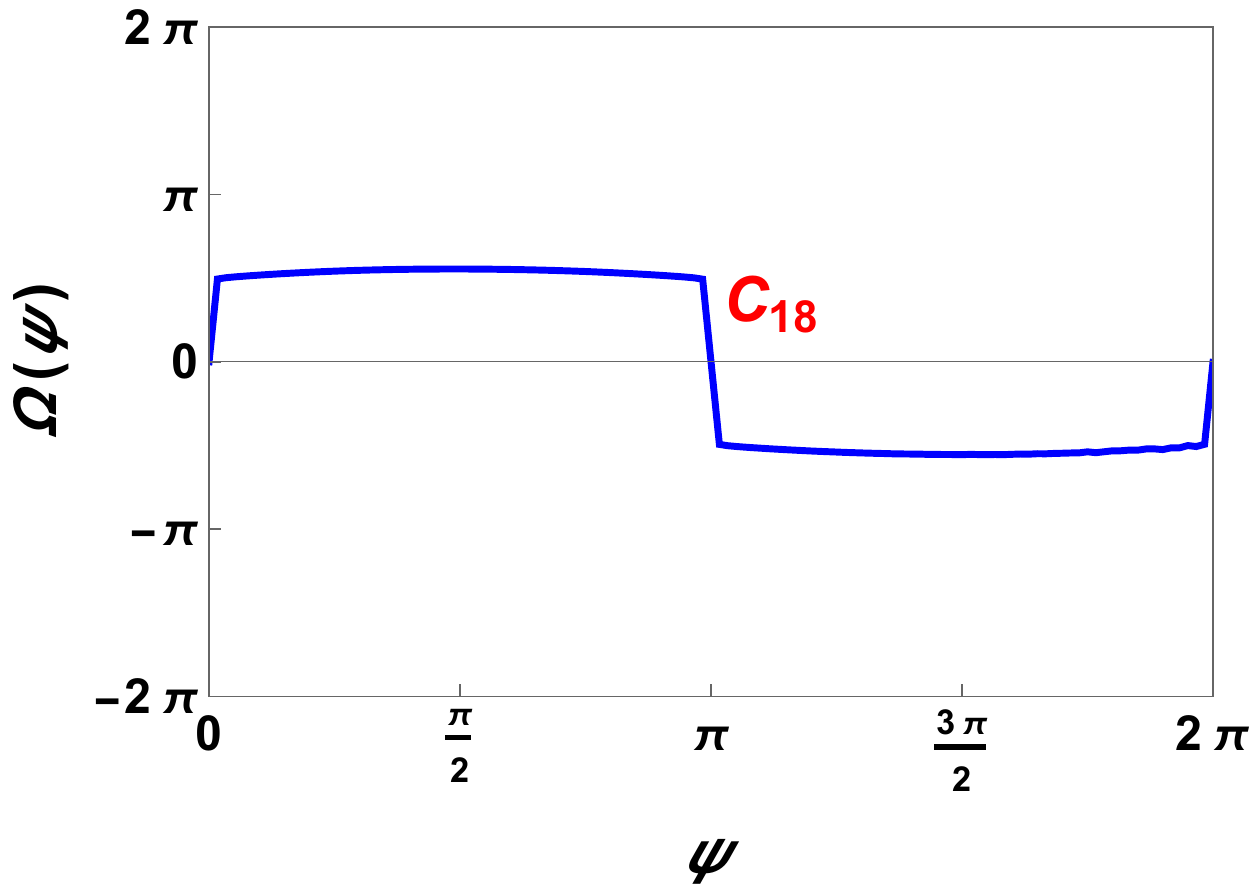}\label{dyC2Amsco}}
    \caption{The second case with $l=l_{ISCO}$ for the dyonic black holes. (a) The unit vector field $n$ on a portion of the $\theta$-$r$ plane. (b) Deflection angle \(\Omega(\psi)\). ``$TP_{13}$" and ``$TP_{14}$" are two TCOs, and ``$MP_{2}$" is the MSCO. The closed loops $C_{16}$, $C_{17}$, and $C_{18}$ have parametric coefficients ($c_0$, $c_1$, $c_2$)=(2.313, 0.32, 0.51), (6.307, 0.3, 0.5), and (25.380, 0.9, 0.4).}
    \label{dycase2}
\end{figure}

\begin{figure}
    \centering
    \subfigure[]{\includegraphics[width=6cm]{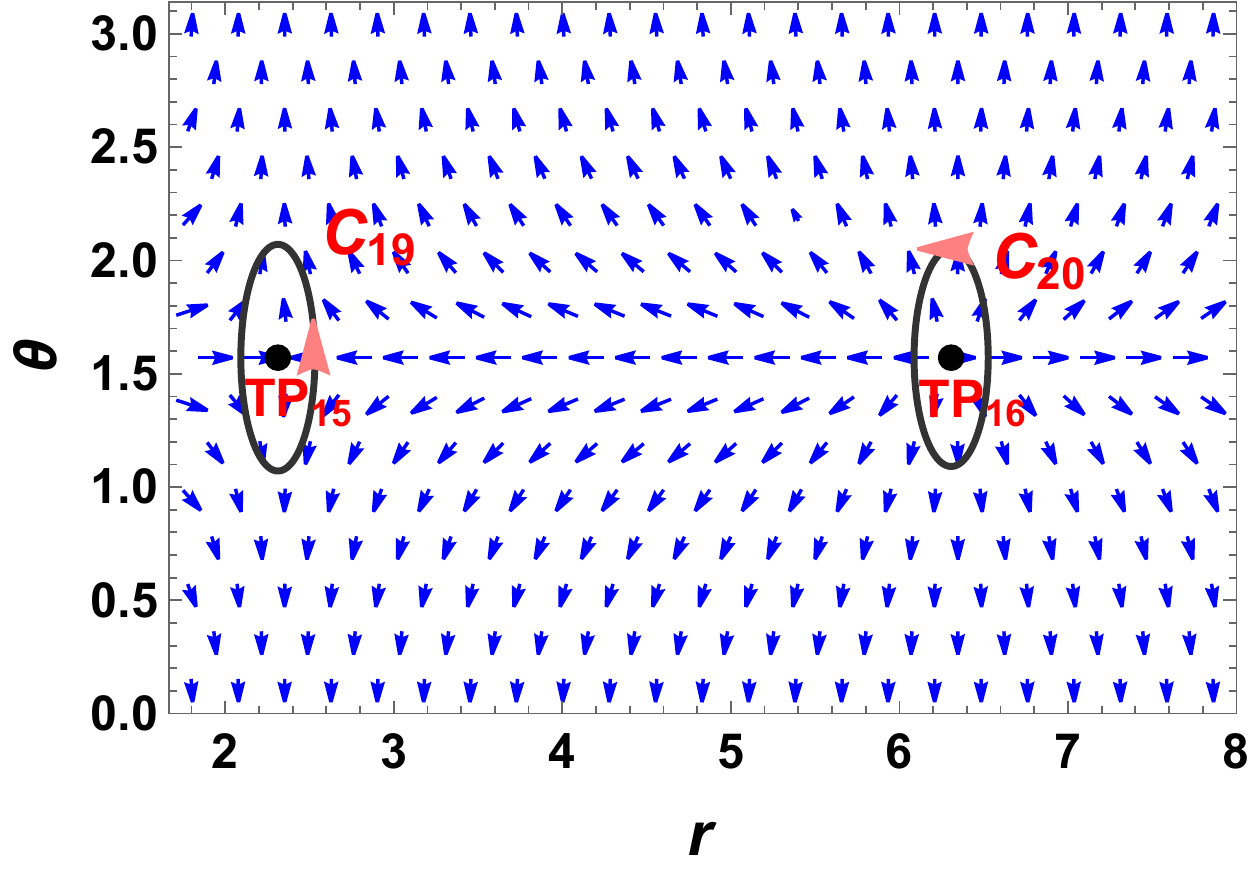}\label{dyC3N12}}
    \subfigure[]{\includegraphics[width=6cm]{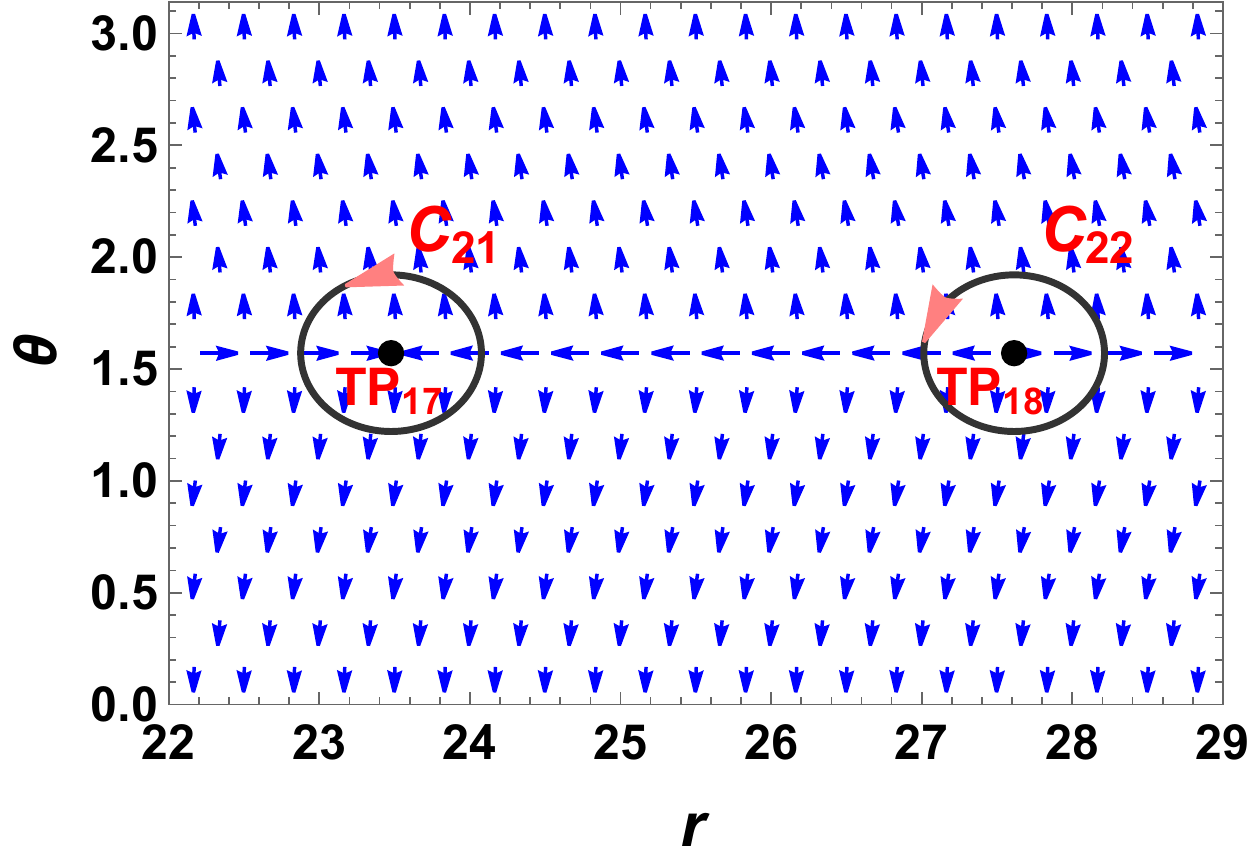}\label{dyC3N34}}
    \subfigure[]{\includegraphics[width=6cm]{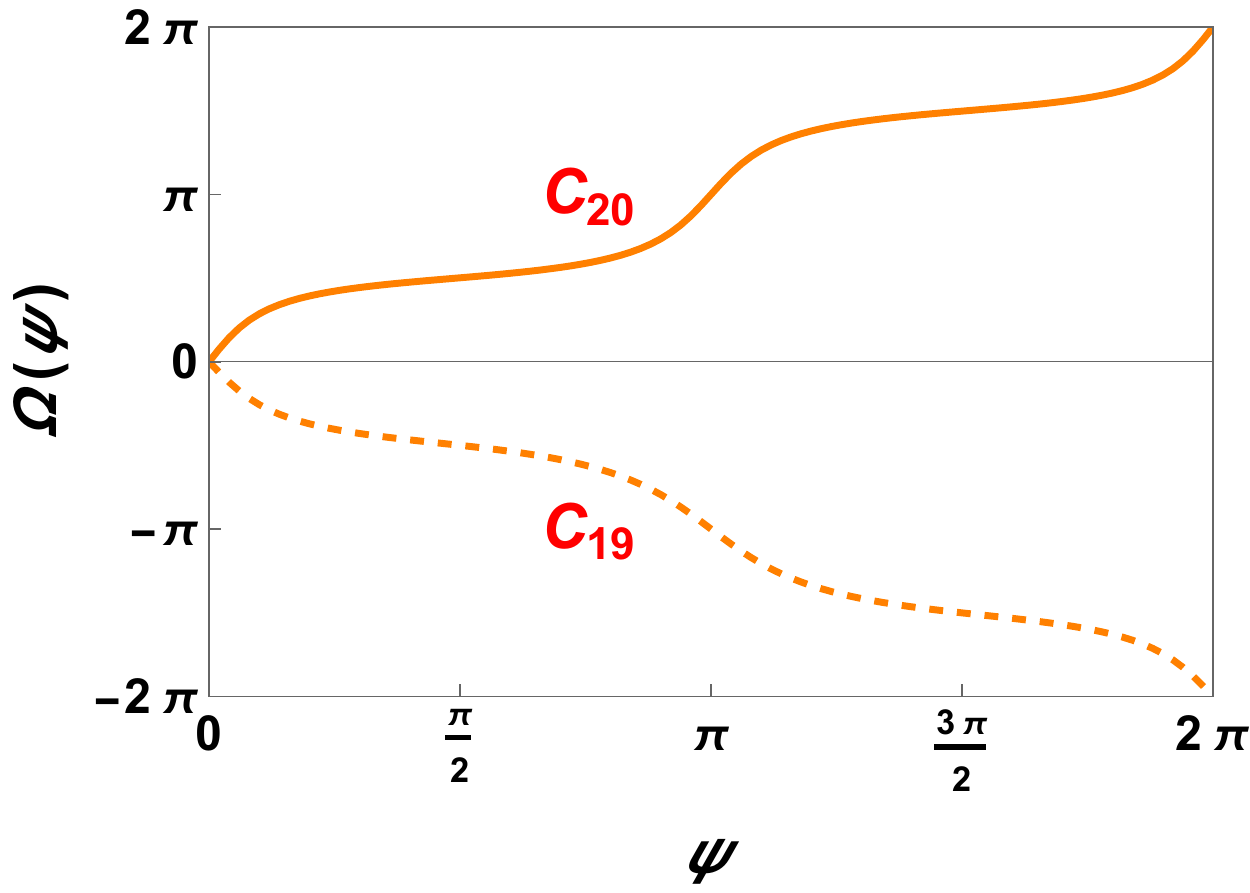}\label{dyC3A12}}
    \subfigure[]{\includegraphics[width=6cm]{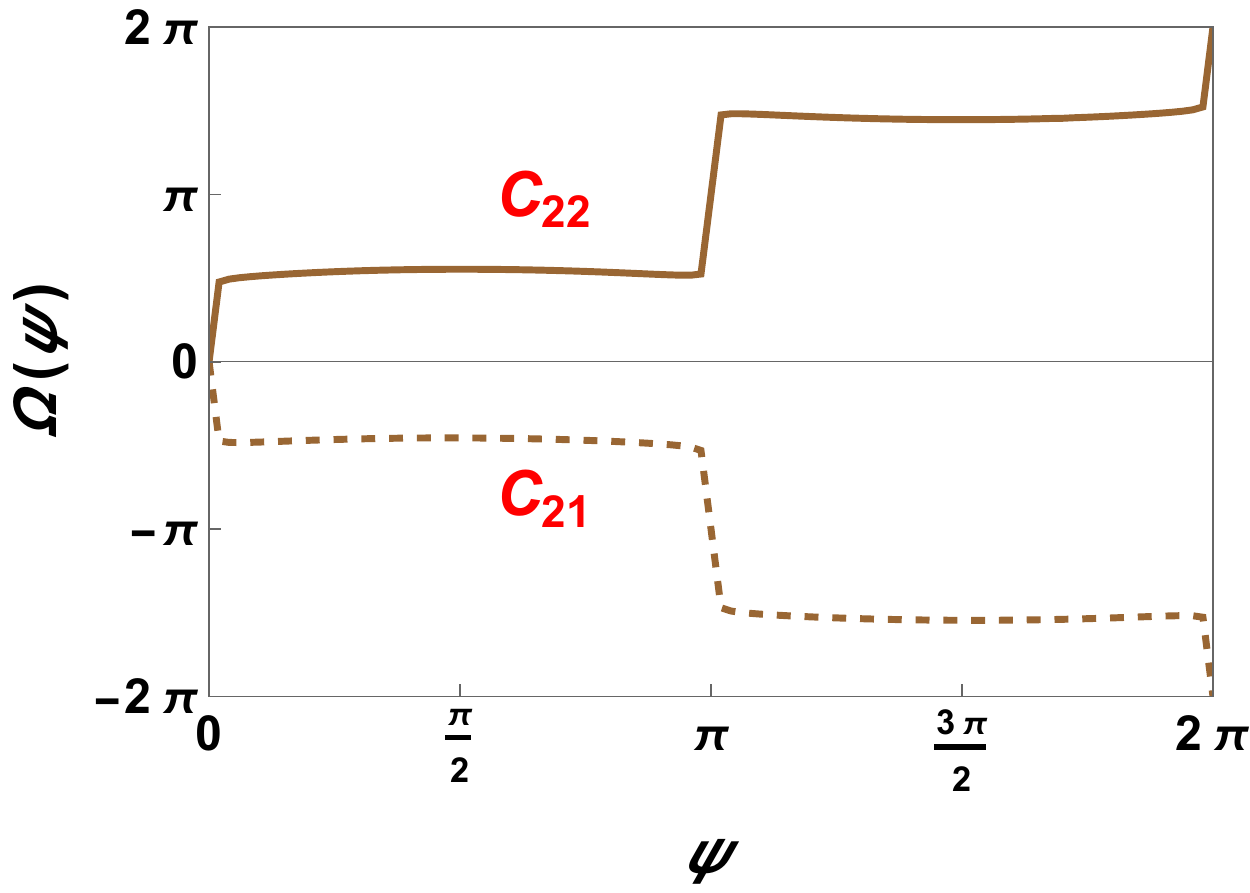}\label{dyC3A34}}
    \caption{The third case with $l=18.58$ for the dyonic black holes. (a) The unit vector field $n$ on a portion of the $\theta$-$r$ plane. (b) Deflection angle \(\Omega(\psi)\). ``$TP_{15-18}$" are four TCOs. The closed loops $C_{19-22}$ have parametric coefficients ($c_0$, $c_1$, $c_2$)=(2.313, 0.22, 0.5), (6.307, 0.22, 0.48), (23.477, 0.6, 0.35), and (27.614, 0.6, 0.35).}
    \label{dycase3}
\end{figure}

Concentrating on the other two cases, we plot the unit vector \(n\) and the deflection angle $\Omega(\psi)$ in Figs. \ref{dycase2} and \ref{dycase3}. When $l=l_{MSCO}$, the vector \(n\) admits three zero points, two of them are the TCOs and the large one is MSCO, see Figs. \ref{dyC2N12} and \ref{dyC2Nmsco}. After counting $\Omega(\psi)$ along three closed loops shown in Fig. \ref{dyC2A12}, we find these two small TCOs have winding numbers -1 and +1, while the MSCO does not attribute to the winding number, see Fig. \ref{dyC2Amsco}. As a result, the topological number $W=-1+1+0=0$ as expected. Taking $l=18.58>l_{MSCO}$ for the third case, we observe four zero points of the vector $n$, which, respectively, locate at $r$= 2.3125, 6.3071, 23.4766 and 27.6136, displayed in Figs. \ref{dyC3N12} and \ref{dyC3N34}. After constructing the closed loops, we find their winding numbers are -1, +1, -1, +1 through the deflection angle $\Omega(\psi)$ given in Figs. \ref{dyC3A12} and \ref{dyC3A34}. Summing them, we obtain the topological number $W=-1+1-1+1=0$, which remarkably indicates that $W$ still does not change for large angular momentum.

\subsection{Topological configuration}

In the previous section, we have examined the topology of the TCOs for the dyonic black holes. With different angular momentum $l_t$, the number of the TCOs changes. So here we turn to study the evolution of the TCOs as a function of $l_t$.

\begin{figure}[h]
    \centering
    \subfigure[]{\includegraphics[width=6cm]{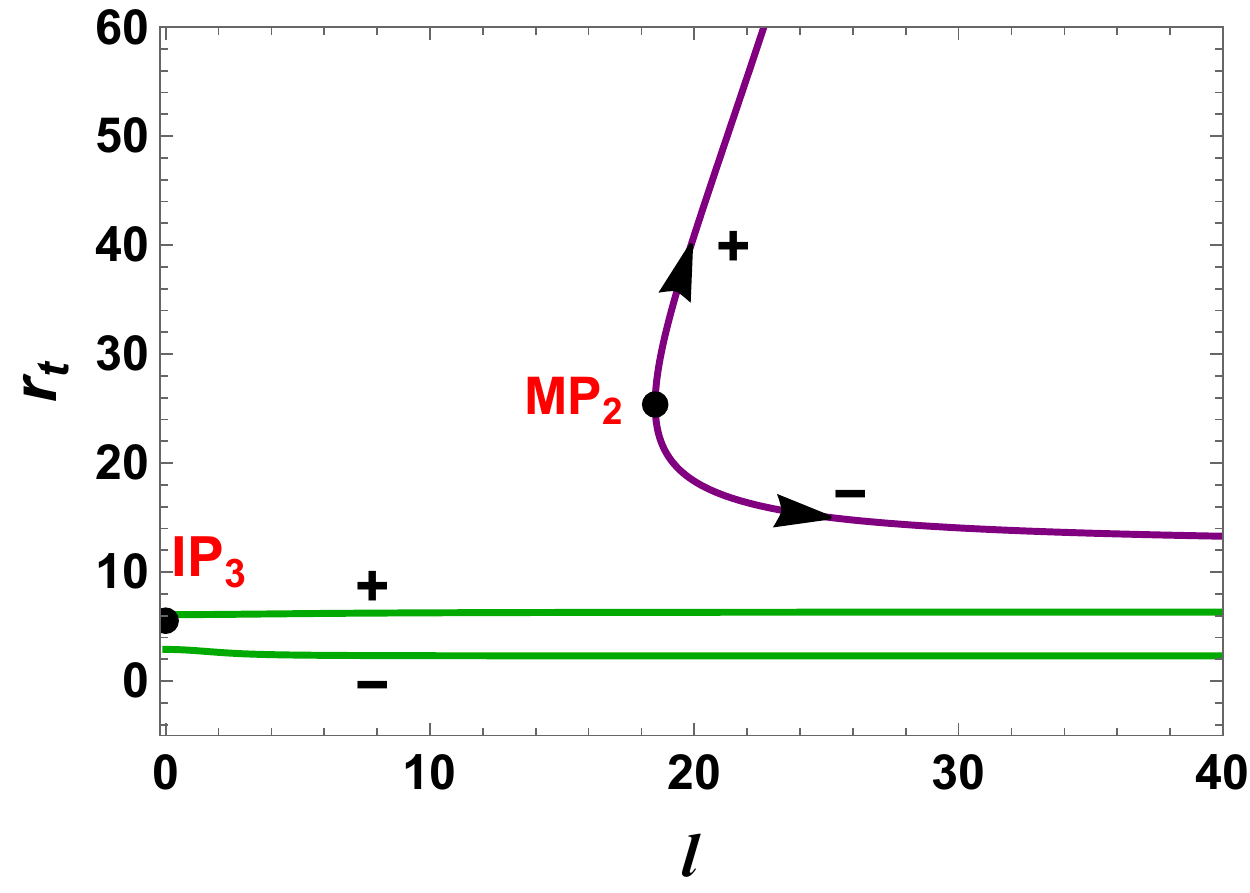}\label{dyRL}}
    \subfigure[]{\includegraphics[width=6cm]{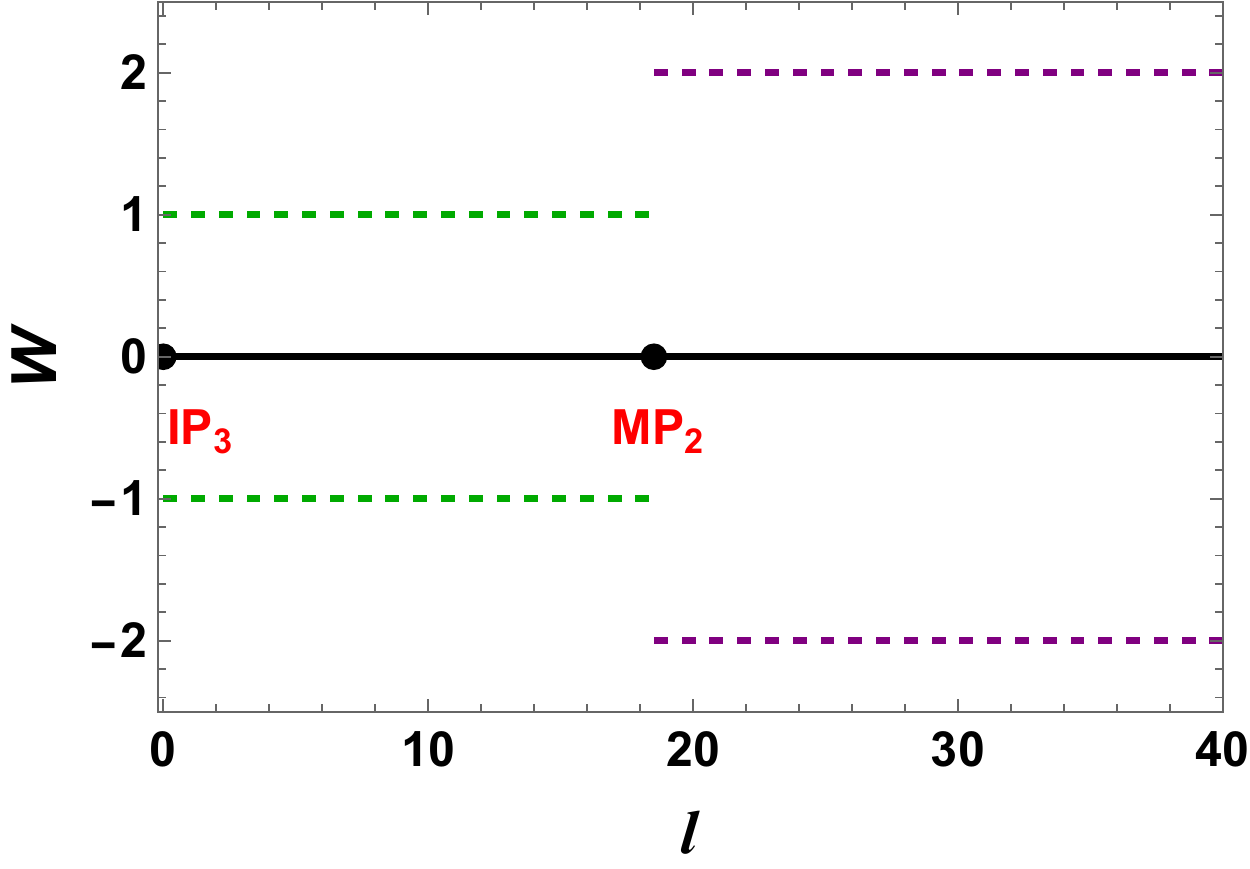}\label{dyWL}}
    \caption{(a) The evolution of TCO radius \(r_t\) vs. the angular momentum \(l\) for the dyonic black holes. The ``$\pm$" denote the positive or negative winding numbers for the TCO branches. (b) The topological number (solid black line) and winding numbers of the TCO branches (dashed green and purple lines).}
\end{figure}

First, we perform the Taylor expansion near $r_{MSCO}$ denoting a bifurcation point,
\begin{equation}
    l_t=l_{MSCO}+ 0.0140657 (r-r_{MSCO})^2+\mathcal{O}\left( (r-r_{MSCO})^3 \right).
\end{equation}
Since $l_t''(r_{ISCO})=0.0140657>0$, the MSCO acts as a generated point. This behavior can also be clearly observed in Fig. \ref{dyRL}. After $l_{MSCO}$, two TCO branches originate. A simple calculation shows that the upper and lower branches have positive and negative winding numbers. Their sum vanishes keeping the same with that of the MSCO.

On the other hand, there are two extra TCOs at small $r$ caused by the quasi-topological electromagnetism term. Both them start at $l=0$ and extend to large $l$ forming two new TCO branches described by green curves in Fig. \ref{dyRL}. Interestingly, they have different values of the winding number.

Such characteristic behavior of the TCOs is significantly different from that of the Schwarzschild black holes and scalarized Einstein-Maxwell black holes. For convenience, we sum the winding numbers for the TCOs in Fig. \ref{dyWL}. For $l<l_{MSCO}$, there is a pair TCOs, respectively having $w$=1 and -1. When the angular momentum is beyond $l_{MSCO}$, a new pair TCOs emerge. However the topological number $W$ always vanishes.

In summary, for a spherically symmetric dyonic black hole, there is a different topological configuration of TCOs from previous black hole solutions. There are two pairs TCOs at most for large angular momentum and one pair TCOs at least for any small angular momentum. However, the topological number still vanishes keeping the same with previous black holes. This result also obviously supports our general result given in Sec. \ref{sec2}.

\section{Conclusions}\label{sec6}

In this work, we studied the topology of the TCOs for the generic spherically symmetric and asymptotic flat black holes. The results suggest that the total topological number of TCOs is zero for each given angular momentum. Then we extended the study to the Schwarzschild black hole, scalarized Einstein-Maxwell black hole, and dyonic black hole. All of them show different characteristic topological configurations. Nevertheless, their topological number is always zero, which indicates they belong to the same topological class.

At first, we considered a static, spherically symmetric, and asymptotic flat black hole. Starting with the Lagrangian of a massive test particle, we obtained the corresponding effective potential, through which the TCOs can be well determined. By making use of the effective potential, we constructed a vector $\phi$ in the $r$-$\theta$ plane with its zero points exactly denoting the TCOs. Employing this property, we established the topology for the TCOs. Each TCO is endowed with a winding number and the stability of the TCOs can be reflected via it. Globally, the topological number $W$ defined as the sum of all the winding numbers can give us information on the number of the TCOs for the black holes. From the asymptotic behaviors of the vector $\phi$, there is a vanishing total topological number for each angular momentum. Consequently, the TCOs always come in pairs. Locally, the stable and unstable TCOs have positive or negative winding numbers, respectively. Meanwhile, the MSCO can be treated as the bifurcation point with vanishing winding number.

Then we generalized the results to three kinds of black holes. For different angular momentum, they show different features of the TCO, and thus have three different topological configurations. For clarity, we summarize them in Table \ref{tab1}.

\begin{table}[h]
    \setlength{\tabcolsep}{3mm}{
    \begin{center}
    \begin{tabular}{cccc}
    \hline\hline
                    & Sch BH   & SEM BH    & dyonic BH \\\hline
    small $l$       & 0        & 0         & one pair  \\
    large $l$       & one pair & one pair  & two pairs \\
    very large $l$  & one pair & two pairs & two pairs \\
    Bifurcation point & 1      & 2         & 1         \\
    \hline\hline
    \end{tabular}
    \caption{Numbers of TCOs and bifurcation points for Schwarzschild black hole (Sch BH), scalarized Einstein-Maxwell black hole (SEM BH), and dyonic black hole. Obviously, they have different configurations of TCOs.}\label{tab1}
    \end{center}}
\end{table}

For the Schwarzschild black holes, we divided them into three cases: $0\leq l<l_{ISCO}$, \(l=l_{ISCO}\), and \(l_{ISCO}<l<\infty\) according to the angular momentum of the ISCO. For different cases, TCO behaves differently. For the first case, TCO does not exist. However, for the third case, two TCOs appear, one of which is stable and the other one is unstable. Nevertheless, the result indicates that $W=0$ is independent of the angular momentum.

Further, we took the scalarized Einstein-Maxwell black hole as an example. Different from the Schwarzschild black hole, the MSCO and ISCO do not coincide due to the presence of the scalar hair, and which gives an interesting topological configuration of the TCOs. With the increase of the angular momentum, we observed that there may be no TCO, one pair TCOs, and two pairs TCOs. Nonetheless, the topological number $W$ still vanishes, keeping the same with that of the Schwarzschild black hole.

As a third example, we considered the dyonic black holes with a quasi-topological term. Due to the quasi-topological term, the pattern of the TCOs is modified. For the previous two cases, the TCO is absent for small angular momentum. However, there will be at least one pair TCOs for the dyonic black hole, providing us with a novel topological configuration. As a result, there will be one pair, two pairs TCOs with the increase of the angular momentum. More interestingly, the innermost and outermost TCOs have $w=-1$ and 1, respectively.

In conclusion, we in this paper considered the topological property and configurations of the TCOs for a spherically symmetric and asymptotic flat black hole. For each angular momentum, the TCOs always come in pairs, which is the same as that of the Kerr black hole. In particular, we exhibit three different topological configurations of the TCOs, see Table \ref{tab1}. Via calculating their topological number, the results state that its value is always zero. So different configurations of TCOs could belong to the same topological class. Therefore, topological information encoding in different configurations of the TCOs for the static black holes is uncovered. As a future study, more interesting topological properties of the TCOs remain to be disclosed, especially in the modified gravity.

\section*{Acknowledgements}
We thank Prof. Peng Wang for sharing their Mathematica file of hairy black holes. This work was supported by the National Natural Science Foundation of China (Grants No. 12075103 and No. 12247101).

\end{document}